\documentclass[useAMS,usenatbib]{mn2e}
\usepackage{graphicx}
\usepackage{amsmath}
\usepackage{amssymb}
\usepackage{float}
\usepackage[font=footnotesize]{caption}
\usepackage{cleveref}
\usepackage{threeparttable}
\usepackage{fixltx2e}

\newcommand{\atlas}{{$\rm{ATLAS}^{3D}$}}

\newcommand{\zsun}{\ensuremath{\rm{Z}_{\odot}}}

\newcommand{\Chi}{$\chi^{2}\!\!_{\mathrm{red }}$}

\title[]{Comparison of Stellar Population Model Predictions Using Optical and Infrared Spectroscopy 
}
\author[C. M. Baldwin et al.]{C.~Baldwin$^{1}$ \thanks{E-mail: christina.baldwin@mq.edu.au},
R.~M.~McDermid$^{1,2}$, H.~Kuntschner$^{3}$, C.~Maraston$^{4}$, C.~Conroy$^{5}$
\\
$^{1}$Department of Physics and Astronomy, Macquarie University, Sydney NSW 2109, Australia \\ $^{2}$Australian Gemini Office, Australian Astronomical Observatory, PO Box
915, Sydney, NSW 1670, Australia \\ $^{3}$European Southern Observatory (ESO), Karl-Schwarzschild-Strasse 2, D-85748 Garching, Germany \\ $^{4}$Institute of Cosmology and Gravitation, University of Portsmouth, Dennis
Sciama Building, Portsmouth PO1 3FX, UK \\ $^{5}$Harvard-Smithsonian Center for Astrophysics, 60 Garden Street, Cambridge, MA 02138, USA }

\begin{document}

\pagerange{\pageref{firstpage}--\pageref{lastpage}} \pubyear{2002}

\maketitle

\label{firstpage}

\begin{abstract}
We present Gemini/GNIRS cross-dispersed near-infrared spectra of 12 nearby early-type galaxies, with the aim of testing commonly used stellar population synthesis models.
We select a subset of galaxies from the \atlas\/ sample which span a wide range of ages (SSP-equivalent ages of 1--15~Gyr) at approximately solar metallicity. 
We derive star formation histories using four different stellar population synthesis  models, namely those of \cite{bruzual_stellar_2003}, 
Conroy, Gunn \& White (2009; 2010), \cite{maraston_stellar_2011} and \cite{vazdekis_uv-extended_2016}. 
We compare star formation histories derived from near-infrared spectra with those derived from optical spectra using the same models. 
We find that while all models agree in the optical, the derived star formation histories vary dramatically from model to model in the near-infrared. We find that this variation is largely driven by the choice of stellar spectral library, 
such that models including high quality spectral libraries 
provide the best fits to the data, and are the most self-consistent when comparing optically-derived properties with near-infrared ones. We also find the impact of age variation in the near-infrared to be subtle, and largely encoded in the shape of the continuum, meaning that the common approach of removing continuum information with a high-order polynomial greatly reduces our ability to constrain ages in the near-infrared.

\end{abstract}

\section{Introduction}

The near-infrared is an important wavelength range which traces populations of a range of ages, suffers lower dust obscuration than the optical (K band extinction is generally more than an order of magnitude
lower than in B; \cite{peletier_extinction_1995}), and 
will be the wavelength range probed by 
facilities like the James Webb Space Telescope (JWST) and adaptive-optics corrected Extremely Large Telescopes (ELTs). In spite of this, 
the near-infrared remains a poorly explored wavelength regime, 
largely due to the difficulty in observing and calibrating ground-based 
near-infrared data, compared to optical data.

In the integrated light of galaxies, the near-infrared is dominated by cool stars, which may be either luminous evolved giants, or slow-burning low-mass dwarfs. 
A particular phase of interest is the 
thermally-pulsing asymptotic giant branch (TP-AGB) phase. TP-AGBs are 
cool, intermediate mass giants, which are thought to dominate the light of stellar populations with ages between $\sim 0.2 - 2$~ Gyr \citep{maraston_evolutionary_2005}. 
TP-AGB stars may contribute up to 40~\% of the bolometric light of populations in this age range, and up to 80~\% in K band \citep{maraston_evidence_2006}.
The complex physics of this phase \citep{marigo_evolution_2008} make it difficult to model theoretically, and its short lifetime makes it difficult to calibrate observationally \citep{maraston_evolutionary_2005}. Subsequently, current 
near-infrared stellar population synthesis models make widely divergent predictions for the age range in which the TP-AGB dominates. 

The Maraston models \citep[][hereafter M98, M05 and M11]{maraston_evolutionary_1998, maraston_evolutionary_2005,maraston_stellar_2011}, for example, which use the fuel consumption theorem of \cite{renzini_energetics_1981}, include a large contribution from stars in the TP-AGB phase compared to traditional models such as \citet[][hereafter referred to as BC03]{bruzual_stellar_2003}, which use the isochrone synthesis technique. 
This large contribution in the Maraston models leads to the presence of strong molecular carbon and oxygen absorption features throughout the near-infrared spectra of intermediate age populations, which could in principle be used as a tracer of these populations. 
These strong absorption features are not found in other popular SPS models such as BC03, which give much less weight to the TP-AGB phase. 

Use of the Maraston models to derive properties of high redshift galaxies leads to estimates of age and mass that are  systematically lower than those derived using the BC03 models (by $\sim 60\%$ on average), highlighting the importance of a correct treatment of this phase \citep{maraston_evidence_2006}.
Since 2005, studies on this topic have produced conflicting results. 
Observationally, the molecular bands predicted by the M05 models have been detected in a number of studies \citep[][and others]{lyubenova_integrated_2012,riffel_stellar_2015}, however, authors were typically unable to strongly link these features to intermediate-age populations. These results are in contrast to the study by \cite{zibetti_near-infrared_2013} who specifically targeted post-starburst galaxies. These galaxies fall in the exact age range at which the TP-AGB phase should dominate, and yet the authors did not observe the strong molecular features predicted by the Maraston models. 
Photometric studies show similar conflicts, with studies providing support both for \citep{maraston_evidence_2006,macarthur_integrated_2010} and against \citep{conroy_propagation_2010,kriek_spectral_2010} a large TP-AGB contribution. Newer models have since been created which use their own prescription for the TP-AGB phase (i.e. the Flexible Stellar Population Synthesis models by \cite{conroy_propagation_2009}), or use new high resolution near-infrared libraries \citep{rock_miles_2016}, adding to the variety of model predictions for this wavelength range.

As well as features relevant to intermediate age populations, 
the near-infrared contains spectral lines which, in old populations, can be used to probe the initial mass function - a fundamental galaxy property which has recently been found to vary systematically in early-type galaxies as a function of stellar mass. Most of the evidence for a systematically varying IMF has come from features blueward of 1~\micron \, (Na\,8200, FeH, CaT, etc.), however features which are differentially sensitive to giant or dwarf stars are found throughout the entire near-infrared range. Some are longer wavelength transitions of well-known IMF tracers (e.g. Na\,I doublets at 1.14~\micron\/ and 2.21~\micron), while others are new (K\,I doublets at 1.17~\micron\/  and 1.25~\micron, CO bandhead at 2.3~\micron).

In this paper, we explore this spectral region and various contemporary models for a well-defined sample of galaxies spanning a range of optical ages (from post-starburst to ancient and passive) to compare the quality of fits of the different models, examine the impact of model choice on the recovered star formation history, and compare the derived properties with those measured at optical wavelengths.
We use high signal-to-noise near-infared spectroscopy of a sample of well-studied galaxies, 
so we are able to compare the properties derived using various near-infrared models with those derived for the same populations using a range of other wavelength regimes, including {\it aperture-matched} optical spectroscopy from the \atlas survey \citep{cappellari_atlas3d_2011}.

The paper is organized as follows. In Section 2 we describe the 
sample. Section 3 details the observations and the data reduction. In Section 4 we give a description of the four models we intend to test, namely, BC03, M11, Flexible Stellar Population Synthesis models \citep[][hereafter FSPS]{conroy_propagation_2009, conroy_propagation_2010}  and the Vazdekis models, which have been recently extended to the near-infrared
by \citet[hereafter V16]{rock_miles_2016}. Our results are found in Sections 5, followed by our discussion and conclusions.

\section{Sample}
\label{sec:sample}

\begin{figure*}
  \centering
  	\includegraphics[width=\textwidth]{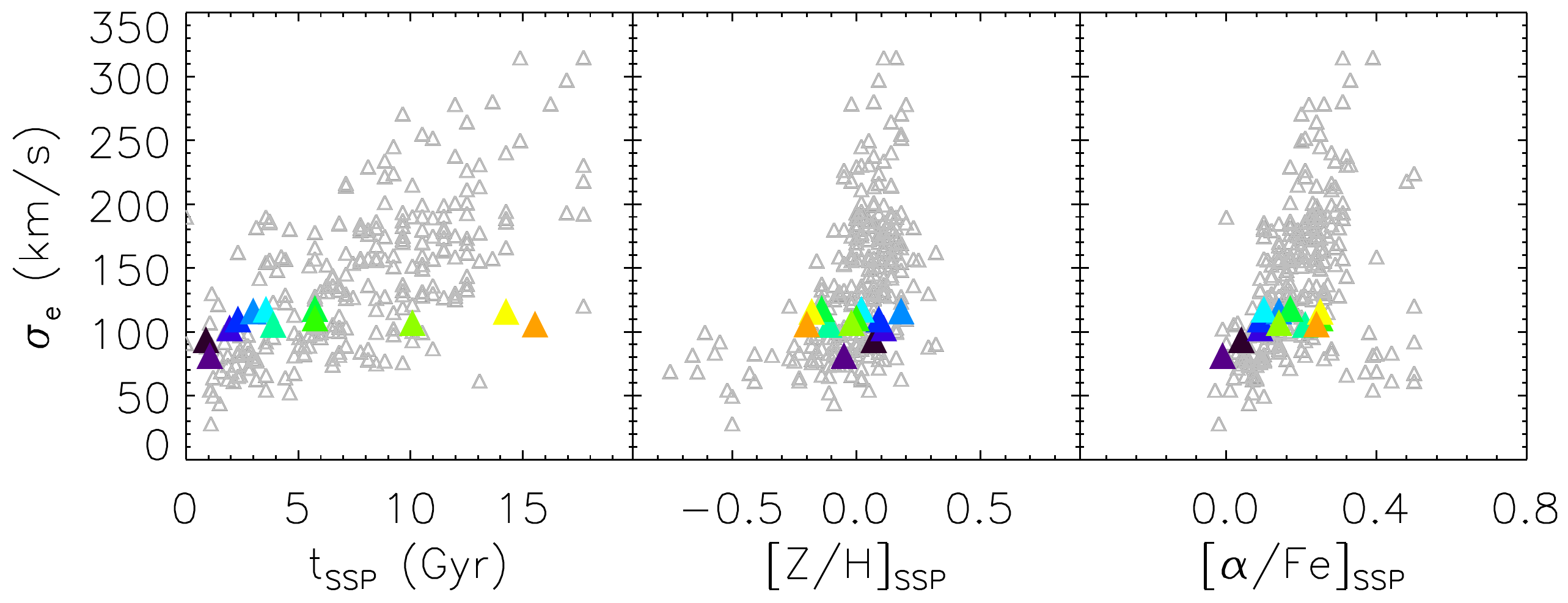}
 	\caption{Sample selection. Panels show (from left to right) mean central stellar age, metallicity and abundance ratio varying against effective stellar velocity dispersion, all derived from \atlas{} optical spectroscopy, calculated in an $R_{\rm e}$/8 aperture. The data and references are given in \Cref{tab:galx}.  Coloured triangles show our sample selection, taking a narrow range in velocity dispersion ($80-120$~km~s$^{-1}$) and a broad range in age. The colours are the same as in \Cref{fig:spec}.}
\label{fig:prop}
\end{figure*}

The early-type galaxies selected for this work were drawn from the \atlas{} survey \citep{cappellari_atlas3d_2011}.
The selection criterion was that the galaxies span a narrow range of velocity dispersion ($\sigma$), as shown in \Cref{fig:prop}. Following the tight scaling relations of stellar populations with $\sigma$ \citep{trager_stellar_2000-2,thomas_epochs_2005,mcdermid_atlas3d_2015}, this narrow range in $\sigma$ ensures that the galaxies are all of approximately solar metallicity and abundances, while spanning a large range of optically-derived SSP-equivalent ages ($1- 15$~Gyr). 
Isolating age as the main variable allows 
us to focus on which features drive the age-sensitivity in this spectral range, and empirically test the impact of the predicted TP-AGB independently of specific models.
The narrow range in velocity dispersion also ensures that intrinsic line width can be compared within the sample without being strongly affected by velocity broadening \citep{kuntschner_line--sight_2004}.

The sample galaxies have excellent supporting data: mean ages, metallicities and $\alpha$-abundances were calculated for all galaxies in the sample by comparing the predictions of the \cite{schiavon_population_2007} single stellar population models to optical Lick indices, and complete star formation histories were also calculated from full spectral fitting of their optical spectra \citep{mcdermid_atlas3d_2015}. 
IC\,0719, NGC\,3032, NGC\,3156, NGC\,3182, NGC\,3489 and NGC\,4710 were shown to contain molecular gas  \citep{combes_molecular_2007, young_atlas3d_2014}, corresponding to regions of ongoing star formation \citep{davis_atlas3d_2014}, and have young mean stellar ages. 
NGC\,3032 and NGC\,3156 are consistent with post-starburst galaxies such as those targeted in \cite{zibetti_near-infrared_2013}.
Other objects show no direct evidence of molecular gas, and have old ages, consistent with having experienced a single burst of star formation at high-redshift.
Our sample therefore covers a very broad range of star formation histories, thus, if there are specific age-dependent features, such as those arising from a short-lived TP-AGB phase, we expect to trace them with this sample.
The basic galaxy parameters are summarised in \Cref{tab:galx}.

\begin{table*}
\begin{threeparttable}
\centering 
\caption{Properties of observed sample.}
\label{tab:galx}
\begin{tabular}{l c  c c c c c c c c} \hline
\multicolumn{10}{c}{\vspace{-0.4cm}} \\ 
Galaxy & Right Ascension & Declination & Velocity & $\sigma_{e}$ & $M_K$ & $t_{ {\rm SSP}}$ & $t_{{\rm m}}$&  [Z/H]$_{ {\rm SSP}}$  & $[\alpha$/Fe]$_{ {\rm SSP}}$ \\  
(1) & (2) & (3) & (4) & (5) & (6) & (7) & (8) & (9) & (10) \\ \hline 

IC0719 & 11 40 18.5 & 09 00 36 & 1833 & 118 & -22.7 & 3.6   & 9.5 & 0.02 & 0.10\\ 
NGC3032 & 09 52 08.1  & 29 14 10  & 1562 & 94 & -22.0 & 0.9 & 2.4  & 0.07 & 0.04\\ 
NGC3098 & 10 02 16.7 &  24 42 40 & 1397 & 118 & -22.7 & 5.7 & 9.8 & -0.14 & 0.17\\ 
NGC3156 & 10 12 41.2 & 03 07 46 & 1338 & 81 & -22.2 & 1.1 &  2.3  &-0.05 & -0.01\\ 
NGC3182 & 10 19 33.0  & 58 12 21 & 2118 & 111 & -23.2 & 5.7 & 10.8 & -0.00 & 0.25\\ 
NGC3301 & 10 36 56.0  & 21 52 56 & 1339 & 110 & -23.3 & 2.3 & 6.3  & 0.09 & 0.09\\ 
NGC3489 & 11 00 18.6  & 13 54 04 & 695 & 103 & -23.0 & 1.9 &  4.4  & 0.11 & 0.09\\ 
NGC4379 &  12 25 14.7   & 15 36 27  & 1074 & 106 & -22.2 & 15.5 & 12.6 & -0.20 & 0.24\\ 
NGC4578 & 12 37 30.5  & 09 33 18 & 2292 & 107 & -22.7 & 10.1 &  11.8 & -0.02 & 0.14\\ 
NGC4608 & 12 41 13.3  & 10 09 20 & 1850 & 116 & -22.9 & 14.3 & 12.7 & -0.18 & 0.25\\ 
NGC4710 & 12 49 38.8 &  15 09 56  & 1102 & 106 & -23.5 & 3.9 &  8.9  & -0.11 & 0.21\\ 
NGC5475 & 14 05 12.4 & 55 44 31 & 1671 & 116 & -22.9 & 3.0 & 7.0 & 0.18 & 0.14\\ \hline 
\end{tabular}
\begin{tablenotes}
\setlength\labelsep{0pt}
\footnotesize
\item \textbf{Columns : }(1) Galaxy name. (2) Right ascension (J2000.0) (3) Declination (J2000.0). (4) Heliocentric velocity in km s$^{-1}$. (5) Effective velocity dispersion $\sigma_e${} measured within 1~$R_{\rm e}$ (km s$^{-1}$). (6) Absolute K-band magnitude. (7) SSP-equivalent age in Gyr. (8) mass-weighted mean age from optical full spectral fitting (Gyr) (9) SSP-equivalent total metallicity.  (10) SSP-equivalent $\alpha$-abundance.  Columns (4)-(6) were taken from Table 3 in \protect \citet[][columns 7,9 respectively]{cappellari_atlas3d_2011}. Columns (6)-(9) were taken from Table 1 in \protect \cite{mcdermid_atlas3d_2015}, and are values calculated on the galaxies' optical spectra within an $R_{\rm e}$/8 aperture. 
\end{tablenotes}
\end{threeparttable}
\end{table*}

\section{Observations and Data Reduction}
\label{ch:obs}
\subsection{Near-Infrared Spectroscopy}

High signal-to-noise ratio near-infrared spectra were obtained over six nights between 1 February, 2012 and 1 May, 2012, using the Gemini Near-Infrared Spectrograph (GNIRS) on the Gemini North 8\,m telescope in Hawaii. They were taken in queue mode through observing program GN-2012A-Q-22, with observing conditions constraints of 85th percentile for seeing (1\farcs1 or better in V-band), 50th percentile for cloud cover (photometric), and unconstrained water vapour and sky background.
GNIRS cross-dispersed (XD) spectroscopy mode was utilised, providing a complete spectrum from 0.8-2.5 $\micron$ across five orders.
Use of the short (0.15''/pixel) blue camera, a 32 l/mm grating, and a slit of width 0.3'' gives an instrumental resolution of $R\sim 1700$.

\begin{table*}
\begin{threeparttable}
\centering 
\caption{Observing Log}
\label{tab:obs}
\begin{tabular}{l c c c c r} \hline 
\multicolumn{6}{c}{\vspace{-0.4cm}} \\ 
Galaxy & Date & N{\scriptsize exp} x T{\scriptsize exp} (s)  & Slit Angle & Airmass & S/N$_K$ \footnote{} \\  
& (dd-mm-yyyy)   &  & $^{\circ}$E of N  & \\ \hline
IC0719 & 04-02-2012 & 8x120 & 30 & 1.1 & 67 \\
NGC3032 & 10-02-2012 & 6x120 & 267 & 1.3 & 78 \\
NGC3098 & 05-02-2012 & 6x120 & 147 &1.0 &  89 \\
NGC3156 & 04-02-2012 & 10x120 & 310 & 1.0 &  101 \\
NGC3182 & 04-02-2012 & 8x120  & 160  & 1.2 & 99 \\
NGC3301 & 04-02-2012 & 4x120 & 265 & 1.0 & 80 \\
NGC3489 & 01-02-2012 & 4x120 & 75 & 1.4  & 198 \\
NGC4379 & 03-02-2012 & 6x120 & 75 & 1.1 & 52 \\
NGC4578 & 04-02-2012 & 6x120  & 50  & 1.1 & 102 \\
NGC4608 & 01-05-2012 & 8x120 & 56 & 1.1 & 92 \\
NGC4710 & 10-02-2012 & 6x120 & 284 & 1.2 &  104 \\
NGC5475 & 10-02-2012 & 7x120 & 222 & 1.4 & 68 \\
\hline 
\end{tabular}
\begin{tablenotes}
\setlength\labelsep{0pt}
\footnotesize
\item $^{1}$ Average signal-to-noise between $2.1-2.2\micron$ calculated as the ratio of the median of the galaxy spectra and the standard deviations of the residuals from the pPXF best fit template. 
\end{tablenotes}
\end{threeparttable}
\end{table*} 
For each observation, the slit was oriented at the mean parallactic angle at the time of observation to minimise the impact of losses due to atmospheric refraction. As the galaxies are extended objects and the slit used in XD mode is short (7''), the telescope was offset 50'' from the object to observe a blank region of sky in order to facilitate the removal of night sky emission lines and dark current Object-sky pairings were observed in an ABA pattern, giving a neighbouring sky frame to every galaxy exposure. 
Immediately after each sequence of exposures, a set of calibration images was taken with the telescope at the position of the galaxy, consisting of continuum and emission-line lamps for flat fielding and wavelength calibration purposes, respectively. This ensured minimal instrument flexure between the science and calibration exposures. 
An observing log is given in \Cref{tab:obs}.

The data were reduced using version 2.0 of the XDGNIRS pipeline made public by Rachel Mason via the Gemini DRForum\footnote{http://drforum.gemini.edu/topic/gnirs-xd-reduction-script/}.
The pipeline calls a variety of tasks, mostly from the Gemini IRAF package \citep{cooke_iraf_2005} to convert the original 2D data into a flux-calibrated 1D spectrum. Firstly the spectra were cleaned of pattern noise caused by the detector controller, using python code available online via the Gemini website\footnote{http://www.gemini.edu/sciops/instruments/gnirs/data-format-and-reduction/cleanir-removing-electronic-pattern-0}.  The code uses the periodicity of the noise to establish a correction, which is subtracted from the frames. Radiation events caused by radioactive lens coatings on the GNIRS short camera were identified by comparison with a clean minimum frame at each nod position, and interpolated over using IRAF's {\it fixpix} task. Cleaned files were divided by a flat field created from a combination of quartz-halogen and infrared flats taken after each science observation.
Subtraction of a sky frame removed night sky emission lines as well as other static artefacts in the detector, such as stable hot pixels. 
The GNIRS detector is tilted with respect to the spectral and spatial axes of each order, so the orders were rectified using daytime pinhole flats. The spectra were then wavelength calibrated using an argon arc frame and 1D spectra were extracted from each order using IRAF task {\it apall}. 
Spectra were extracted within an aperture of 
$\pm R_{\rm e}$/8, although in one galaxy (NGC\,4710) the slit length did not allow this and we extracted a spectrum using the full slit length instead ($\pm R_{\rm e}$/12). 
The extraction was carried out such that the sum was weighted by the signal to noise in each pixel \citep{horne_optimal_1986} and the noise statistics of the CCD were used to detect deviant pixels, which were cleaned and and replaced.

An airmass-matched A-type star of magnitude m$_{\rm K} = 6.5 - 8$ was observed before or after each science object to correct for telluric absorption and to approximately flux calibrate the science data. Before applying the telluric correction, the star's intrinsic hydrogen lines were removed by interactively shifting and scaling a model spectrum of the star Vega \citep{kurucz_atlas9_1993}.
IRAF's \textit{telluric} task was then used to interactively shift and scale the (hydrogen line-free) telluric star spectra to achieve the optimal telluric removal in the science target. 

\begin{figure*}
   \rotatebox{90}{
     \begin{minipage}{\textheight}
  	\includegraphics[width=0.93\linewidth]{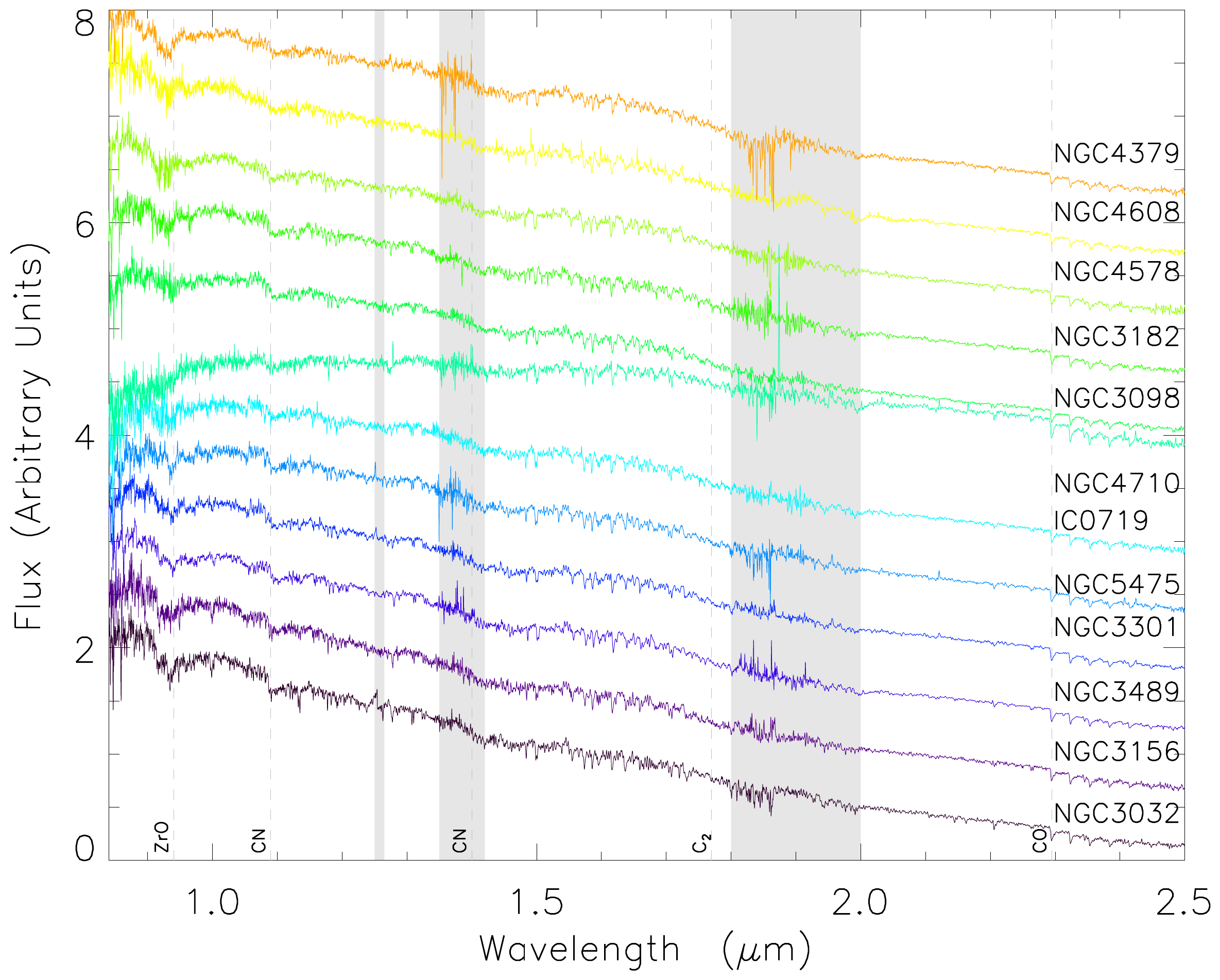}  	
 	\caption{Rest frame spectra of the galaxy sample, sorted by increasing \atlas\/ derived SSP age (age increases upwards). The grey shaded boxes mark regions of poor atmospheric transmission as defined in the text. Important molecular features are marked with dashed lines.}
 \label{fig:spec}
     \end{minipage}
  }
\end{figure*}

For the purposes of flux calibration, the telluric star was approximated as a black-body continuum having the effective temperature of the star’s spectral type \citep{pecaut_intrinsic_2013}. 
This black-body spectrum was scaled to the value of the star's K band magnitude as taken from 2MASS \citep{skrutskie_two_2006}.
Each order of the science target was multiplied by this function. Finally, the orders were joined together using the IRAF task {\it odcombine}, which ensures a smooth overlap region between the different orders.
The resulting spectra\footnote{available on github https://github.com/cbaldwin1/Reduced-GNIRS-Spectra} are shown in \Cref{fig:spec}, ordered by the SSP-equivalent age derived by \atlas.

\subsection{Optical Spectroscopy}
\begin{figure}
  \centering
  	\includegraphics[width=\columnwidth]{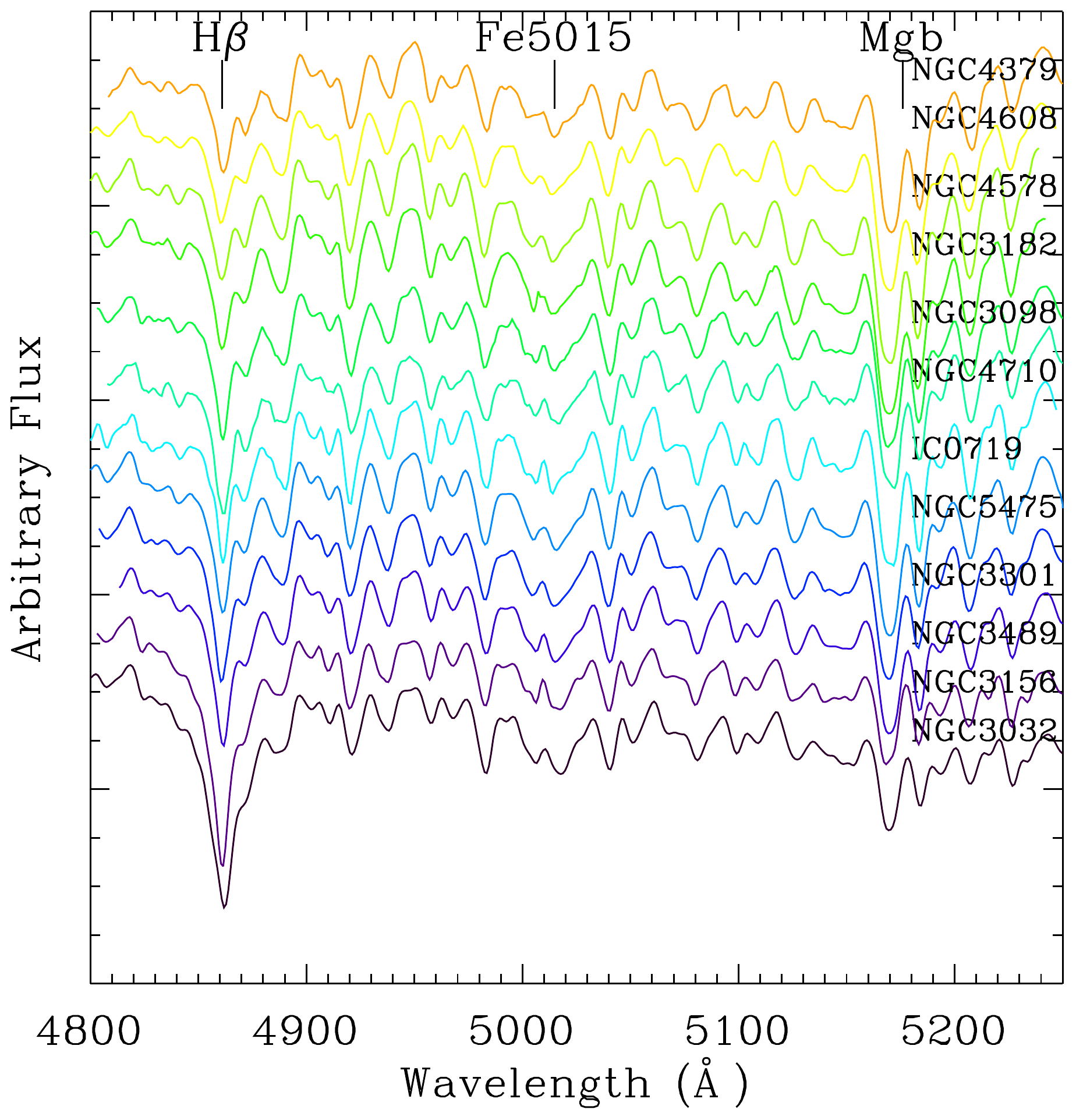}
 	\caption{Rest frame SAURON optical spectra, ordered by decreasing SSP-equivalent age from top to bottom. The colours are the same as in \Cref{fig:spec}. }
\label{fig:opt}
\end{figure}

We also utilise the optical spectroscopy obtained as part of the \atlas survey in order to test the various models for self-consistency.  SAURON spectra cover the wavelength range $4800-5380$~\AA\/ at a spectral resolution of 4.2~\AA\/ full width at half-maximum (FWHM). 
Observations are described in \cite{cappellari_atlas3d_2011} and  the data reduction is detailed in \cite{bacon_sauron_2001}.
To test for self-consistency, we derived star formation histories for all models using both wavelength ranges. 
To accurately compare properties derived using our new GNIRS near-infrared spectroscopy with those derived from SAURON optical integral-field spectroscopy, we re-extract optical spectra integrated along an effective slit of width and length equal to that of the GNIRS aperture, centred on the galaxy nuclei and aligned along the parallactic angle. The optically derived parameters were thus obtained on spectra extracted from the same region as covered by our near-infrared data.
The spectra are shown in \Cref{fig:opt}, ordered by the SSP-equivalent age derived by \atlas.

\section{Stellar Population Synthesis}
\subsection{Stellar Population Models}
There are nowadays a large number of stellar population models available for use in extragalactic work. These models differ in their choice of stellar tracks and spectral libraries, as well as the parameter space they cover (age, metallicity, wavelength range, initial mass function) and their methods used for including various post-Main Sequence phases (e.g. blue horizontal branch, TP-AGB). 
Here we describe the salient features of the models used in this paper.

\subsection*{Maraston \& Stromback (2011)}

The M11 models adopt the same stellar evolution pattern as in \cite{maraston_evolutionary_2005}, namely, they use the isochrones and stellar tracks from \cite{cassisi_intermediate-age_1997} up to the main sequence turn-off, after which time they use the fuel consumption theorem of \cite{renzini_energetics_1981} to calculate the energetics of post-Main Sequence  (PMS) stages of stellar evolution such as the TP-AGB phase. 
With this approach, the luminosity of PMS stars is proportional to the amount of nuclear fuel burned in each PMS phase, making the contribution of each phase independent of the details of stellar evolution. 
The specific contribution of the TP-AGB phase at each age was calibrated against Magellanic Cloud clusters. 

For the stellar spectra,  the M11 models offer the option of four different libraries of flux-calibrated empirical stellar spectra, namely Pickles \citep{pickles_stellar_1998}, ELODIE \citep{prugniel_new_2007}, STELIB \citep{le_borgne_stelib:_2003} and MILES \citep{sanchez-blazquez_medium-resolution_2006, cenarro_medium-resolution_2007}, as well as the MARCS \citep{gustafsson_grid_2008} library of very high resolution theoretical stellar spectra. Each library is supplemented with empirical stellar spectra of O- and C-rich TP-AGB stars from \cite{lancon_modelling_2002}. 
We used the Pickles library to fit the near-infrared data, as it is the only library in the M11 model set to cover the near-infrared at the intermediate ages which are of interest here. The library covers a wavelength range of 1150\,\AA\/ to $2.5~ \mu$m at a resolving power $\rm{R} \approx 500$. Using this library, model SEDs are available up to 25,000 \AA \, for 58 ages, logarithmically spaced from 2.5 Myr to 15 Gyr at solar metallicity only, for three choices of IMF (Salpeter, Chabrier and Kroupa). 
It should be noted that approximately half of the stars in the Pickles library lack spectroscopic observations above 1~\micron. 
At these wavelengths, the ``spectra'' of these stars consist of a 
smooth energy distribution formed from broad-band photometry, 
hence some near-infrared absorption features may be poorly resolved.  For the optical fits, we used the MILES library, which covers $3500-7430$~\AA\/ at a resolution of 2.51~\AA~(FWHM), which is the same optical library used by the majority of other models tested.

\subsection*{Bruzual \& Charlot (2003)}
The BC03 models use the Padova 1994 tracks \citep{alongi_evolutionary_1993, bressan_evolutionary_1993, fagotto_evolutionary_1994} supplemented with the \cite{vassiliadis_evolution_1993} models for TP-AGB
stars and the \cite{vassiliadis_post-asymptotic_1994} models for post-AGB stars.
The default stellar spectral libraries are the empirical STELIB library for the optical \citep[$\rm{R} \approx$ 2000;][]{le_borgne_stelib:_2003}, extended with the semi-empirical BaSeL library \citep[$\rm{R} = 200-500$;][]{westera_standard_2002} for ultraviolet and  near-infrared predictions. 
BC03 also offer the option of generating solar metallicity SEDs extending to the near-infrared using the Pickles library, which has a resolution of $R \sim 500$. This is higher resolution than the BaSeL library, and corresponds to the same library used by the M11 models (see above). 
We therefore use STELIB to fit the optical data, and Pickles to fit the near-infrared. Model predictions are available for the same age and metallicity ranges as the M11 models described above, and for Salpeter and Chabrier IMFs.

\subsection*{Conroy (2009)}

The FSPS models by \cite{conroy_propagation_2009,conroy_propagation_2010} use stellar evolution tracks from the Padova group \citep{marigo_evolution_2007, marigo_evolution_2008}. 
The spectral libraries are MILES in the optical, extended with the semi-empirical BaSeL3.1 library \citep{lejeune_standard_1997, lejeune_standard_1998, westera_standard_2002}, supplemented with empirical TP-AGB spectra from the library of \cite{lancon_modelling_2002}. 
These models introduce parameters $\Delta L$ and $\Delta T$, which allow the user to modify the location of the TP-AGB track on the HR diagram. 
Originally these shifts were with respect to the Padova tracks, however the current models set the default values of $L$ and $T$ to the best-fit values calculated in 
\cite{conroy_propagation_2010}. We used the default model values for this study. 
This model set provides templates for 188 ages from 0.3~Myr -- 15~Gyr and 22 metallicities at the resolution of the BaSeL spectral library ($\rm{R} \approx 200$) 
Models can be computed for a range of standard as well as non-standard IMFs (Salpeter, Chabrier, Kroupa, piece-wise power-law IMF, top-heavy IMF).
The FSPS models allow multiple aspects of the model creation to be modified, however for the purpose of comparison, we use these models in their default state. Unless explicitly stated otherwise, from now on `FSPS' refers to the model set described above.

\subsection*{Vazdekis (2016)}  
The current Vazdekis models (hereafter V16) incorporate a number of recent updates \citep{vazdekis_evolutionary_2010,vazdekis_miuscat:_2012,
vazdekis_evolutionary_2015,rock_stellar_2015,rock_miles_2016}, which extend the original Vazdekis models (2003) from 1,680--50,000 \AA.
Models are available for two sets of stellar tracks:
\citet[hereafter BaSTI]{pietrinferni_large_2004} and \citet{girardi_evolutionary_2000}.
We utilise the  BaSTI tracks, which use the synthetic AGB technique to cover the full AGB phase, including up to the end of the thermal pulses.
The spectral libraries used in the optical and near-infrared ranges are MILES and IRTF \citep{rayner_infrared_2009} respectively.
IRTF is an empirical spectral library covering the J, H and K bands (9500--24,000~\AA) at medium resolution (R $\sim2000$). The library contains $\sim 200$ stars, mostly of solar metallicity, including late-type stars, AGB, carbon and S stars.
Models are available for ages 1--15~Gyr only, for metallicities between 
-0.40 and 0.26, for unimodal and bimodal IMFs of slopes between 0.3 and 3.3.

\subsection*{Model summary}

\Cref{tab:libs} summarises the default basis set we have chosen for each group of models. The basis sets were chosen so as to cover approximately the same parameter space for all models, as well as span the range of properties derived for our sample using optical spectroscopy. As the IRTF spectral library can generate models only for $t\geq1$~Gyr, and the Pickles library for solar metallicity only, this is the range we use for all models. We use Salpeter IMF for all fits.

\begin{table*}
\centering 
\caption{Overview of the default basis sets for each stellar population model used in this study.} 
\begin{tabular}{l c c c c c c c c c c c} \hline
\multicolumn{12}{c}{\vspace{-0.4cm}} \\ 
SPS  & Stellar &  Optical & Optical & Near-Infrared  & Near-infrared & Metallicities & IMF & N$_{{\rm ages}}$ & Ages (Gyr) &  & \\  
Model  & Tracks & Library & Resolution & Library &  Resolution & ($\text{Z}_{\odot}$) & &  &  &  & \\  \hline

M11 & Cassisi  & MILES & 2.5~\AA & Pickles &    500 &  0.02 & Salpeter & 15 & 1--15 &  &  \\
BC03 & Padova (1994) & STELIB & 3~\AA & Pickles &       500 &  0.02 & Salpeter & 15 & 1--15 &  & \\
FSPS & Padova (2008) & MILES & 2.5~\AA  & BaSeL & 200  & 0.02 & Salpeter & 15 & 1--14.96 &  & \\
V16 & BaSTI & MILES &  2.5~\AA  & IRTF & 2000 & 0.02 & Salpeter & 14 & 1--14 &  &  \\
 \hline 
\end{tabular}
\label{tab:libs}
\end{table*} 

\subsection{Star formation histories from full spectral fitting}

We use the penalised pixel fitting code (pPXF)\footnote{http://purl.org/cappellari/idl} by \cite{cappellari_parametric_2004} to fit a linear combination of basis spectra to the data for each model set. The code fits the entire input spectrum, with the exception of regions strongly affected by telluric absorption, which are masked out. 
We note that, while there are many different spectral fitting codes and techniques available (e.g. STECKMAP/Ocvirk, MOPED/Heavens, NBURSTS/Chilingarian, STARTLIGHT/Cid Fernandes, FIREFLY/Wilkinson), cross-comparison of these various techniques have been shown to yield rather consistent results  \cite[e.g.][]{koleva_spectroscopic_2008,ferre-mateu_young_2012,wilkinson_p-manga:_2015, mentz_abundance_2016, maksym_deep_2014}. In particular, \cite{goddard_sdss-iv_2017} explore the effect of changing fitting code and models, using Firefly and STARLIGHT, and conclude that in order to understand population model effects one should keep the fitting approach the same. 
We therefore keep the fitting approach fixed, in order to investigate the differences introduced by the choice of models, where detailed comparisons are less well documented.

pPXF convolves the model templates with a Gaussian line-of-sight velocity distribution, which broadens the model SSPs in order to best fit the data. The convolved templates are then fit to the data as the code searches for the minimum of the equation
\begin{equation}
\chi^2=\sum_{n=1}^N    \frac {(M-O)^2} {\sigma^2}
\label{eq:chi}
\end{equation}
where M is the model spectrum, O is the observed spectrum, and $\sigma$ is the error spectrum. The fractional weights assigned to each template give the mass-weighted contribution of that population to the observed spectrum.

We use regularisation, which is a standard way to solve ill-posed problems \citep{press_numerical_1992}. This is implemented in pPXF via the REGUL keyword, which penalises the $\chi^2$ value when neighbouring weights do not vary smoothly. 
A larger REGUL value leads to a smoother output, by effectively reducing the error associated with the linear regularisation constraints. We use a fixed regularisation value of 50, which is consistent with the typical weights returned by pPXF. 
This gives a unique solution, i.e. it returns the smoothest solution out of the many degenerate solutions that are equally consistent with the data.
The mass-weighted mean age of each galaxy is calculated as the average of the template ages weighted by the template weights, and does not depend strongly on the chosen regularisation value.

\subsubsection{Preparatory Steps}

Prior to the fit, models were resampled onto pixels of constant linear wavelength where necessary. To account for differences in resolution between data and models, we convolved them using a Gaussian of FWHM given by:
\begin{equation}
\rm{FWHM}_{Gaussian} = \sqrt{(\rm{FWHM}_{max}^{2}-\rm{FWHM}_{min}^{2})}
\label{eq:fwhm}
\end{equation}
In every case except V16, the data were higher resolution than the models, and had to be degraded to carry out the fit. 
The observed galaxy spectrum and the model of interest were then rebinned to a log scale so as to have a constant velocity step. 
The log rebinning was done first on the model/data with the coarser velocity step size and then the data/model was rebinned to the same velocity step. As such, the size of the velocity step varied for the different data-model combinations (from $\sim 45$~km/s for V16 and BC, for which the step size was set by the data, 98.5~km/s for M11 and 787~km/s for FSPS).
Regions of poor atmospheric transmission between the J H \& K bands were masked\footnote{Poor atmospheric transmission is defined as where the atmospheric transmission model of \cite{lord_atran_1992} is $\leq50\%$, assuming a water vapour column of 1.6mm and an airmass of 1.0.}, as well as other regions where telluric removal was particularly difficult. 

An initial fit was carried out using all templates simultaneously, including high order additive polynomials. Additive polynomials change the strengths of absorption features, so are not recommended for stellar population analyses, but are useful at reducing template mismatch when deriving kinematics. 
We thus obtained the best numerical fit possible and secured the correct velocity and velocity dispersion values, which were then fixed in the subsequent fits, reducing the freedom pPXF has to compensate the template choice with varying kinematic parameters.

\subsubsection{Treatment of the Continuum }

There are two common strategies when carrying out spectral fitting: one uses a high order polynomial to fit the continuum, such that derived properties are based primarily on the strengths of strong, narrow absorption features in the spectrum. The other is to preserve the continuum, taking advantage of the extra information it encodes. 

We implemented both approaches to determine the effect of each on the derived star formation histories (SFHs). In theory, keeping the continuum shape provides more information on the stellar populations present in the galaxies, however it assumes that both models and data are accurately flux calibrated, and that extinction is understood. In this case, we use the well-known extinction curve of Calzetti \citep{calzetti_dust_2000}, where dust is treated as a simple foreground screen. This option is implemented in pPXF, and characterised by an output colour excess $E(B-V)$. 

We also tested the effect of including a multiplicative polynomial. 
Multiplicative polynomials compensate for large-scale model and data deficiencies driving the fit without affecting the relative strength of absorption lines. We use order 10, as this was found to remove large-scale discrepancies without fitting intrinsic galaxy features. 

\subsubsection{Noise estimate and $\chi2$}
\label{sec:noise}

The $\chi2$ calculated by pPXF depends on an estimate of the noise, as seen in \Cref{eq:chi}. To be able to meaningfully rank the models, all of which are rebinned and interpolated in different ways, is not straightforward. Initially we calculated the noise in each galaxy as the standard deviation of the residuals obtained from a good fit. 
However, the $\chi2$ calculated by pPXF was found to depend on the log rebinning and resolution correction convolution, as both of these processes introduce covariance between pixels, causing the noise estimate to be underestimated for particular models. 

As these corrections are model dependent, we estimated the noise independently for every model set. As before, we carried out a good quality fit using the model with the highest quality near-infrared spectral library (V16) and high order polynomials to ensure a good fit to all the intrinsic galaxy features seen in the data. However, we matched the resolution and velocity step size used in the fit to that of the model of interest, giving a unique noise estimate for every model set, and $\chi2$ values that can be compared between models.

\subsubsection{Age estimate uncertainties }
We used a Monte Carlo approach to estimate uncertainties in the derived mean ages. Random noise was added to the galaxy spectrum which was consistent with the derived error spectrum. 500 realisations were generated for each object, and we again performed full spectral fitting on each of these new realisations, without regularisation. The error in the mean ages was taken to be the standard deviation of the output age array, divided by $\sqrt{2}$, to account for the fact we added noise onto an already noisy spectrum.
This does not take into account systematic errors in the models themselves.

\subsection{Line strengths}
The measurement of line strength indices, when compared with the predictions of SSP templates,  provides an alternative comparison of different stellar population models, which is largely insensitive to uncertainties in flux calibration. To measure the strength of atomic absorption features we follow a procedure similar to that used for the measurement of the well-known Lick indices in the optical \citep{worthey_old_1994}. 
Line strength indices are typically defined by a central bandpass covering the feature of interest, flanked by two pseudocontinuum bandpasses, one to the red side of the feature and one to the blue. The strength of the feature is then calculated as an equivalent width: 
\begin{equation} \label{eq:ew}
\rm{EW} = \int_{\lambda_1}^{\lambda_2} \left(1-\frac{S(\lambda)}{C(\lambda)} \right) d\lambda
\end{equation}
where S($\lambda$) is the observed spectrum and C($\lambda$) is the local pseudocontinuum, which is given by the linear function connecting the average flux in the pseudocontinuum bands.
For one sided molecular `break' features such as CN and C$_2$, we calculate index strength as in \cite{gonneau_carbon_2016}:
\begin{equation}
I(X) = -2.5~ \rm{log_{10}} [F _b (X)/F _c (X)],
\end{equation}
where $F_b (X)$ and $F_c (X)$ are the fluxes in the absorption band region and the pseudo-continuum region of index X.

The two main contributions to the statistical uncertainties in measuring line indices come from the statistical Poisson and detector noise in the index pass bands, and from errors in the measured radial velocity.
The errors on the strengths of the features were calculated using Monte Carlo simulations, by shifting the bandpasses randomly according to the velocity error obtained from pPXF, after adding random noise at a level consistent with the observed error spectrum.

The bandpasses and pseudo-continua we use for each index are listed in \Cref{tab:defs}. The models and data each have a different native resolution, thus, we degrade to the lowest common resolution before measuring indices.
We did not correct for the effect of differing velocity dispersion broadening, as, compared to the difference in instrumental resolution, the difference in index strengths due to $\sigma$ was negligible.

\begin{table*}
\begin{threeparttable}
\centering 
\caption{Index Definitions.}
\label{tab:defs}
\begin{tabular}{l c c c c c r} \hline
\multicolumn{7}{c}{\vspace{-0.3cm}} \\ 
Index & Blue Pseudocontinuum & Central Bandpass & Red Pseudocontinuum$^a$ & Units & Source$^b$ & Notes \\ 
& ($\mu$m)  & ($\mu$m)  & ($\mu$m) & \\ \hline
CN 1.1 & 1.0770 - 1.0830  &  1.0970 - 1.1030 && mag & 1 &  ratio\\ 
Na 1.14 & 1.1337 - 1.1367 & 1.1367 - 1.1420 & 1.1420 - 1.1450 &\AA & 2 \\
CN 1.4 &  1.3800 - 1.3900 & 1.4130 - 1.4230 && mag & 3 &  ratio\\ 
C$_2$ 1.8 &  1.7520 - 1.7620 & 1.7680 - 1.7820&& mag & 1 & ratio \\ 
Na 2.21& 2.1934 - 2.1996 &  2.2000 - 2.2140  &  2.2150 - 2.2190 &\AA & 4 \\
Fe\,a 2.2 & 2.2133 - 2.2176 & 2.2250 - 2.2299 & 2.2437 - 2.2497 & \AA & 5\\
Fe\,b 2.2 & 2.2133 - 2.2176 & 2.2368 - 2.2414 & 2.2437 - 2.24971 & \AA &5\\

Ca 2.2 & 2.2450 - 2.2560 & 2.2577 - 2.2692 & 2.2700 - 2.2720 & \AA & 5 \\ 
Mg 2.2 &  2.2700 - 2.2720 & 2.2795 - 2.2845 & 2.2850 - 2.2874 & \AA & 5 \\
CO 2.3 & 2.2460 - 2.2550 & 2.2880 - 2.3010 & 2.2710 - 2.2770 & mag & 6 \\ 
\hline 
\end{tabular}
\begin{tablenotes}
\setlength\labelsep{0pt}
\footnotesize
\item 
$^a$In the case of CO, `Red Pseudocontinuum' indicates the most redward pseudocontinuum band, which is still blueward of the feature itself.
$^b$\textbf{Sources.} (1) \cite{gonneau_carbon_2016} ; (2) \cite{smith_imf-sensitive_2015} ; (3) newly defined in this paper; (4) \cite{riffel_near-infrared_2011} ; (5) \cite{silva_new_2008} ; (6) \cite{marmol-queralto_new_2008} 
\end{tablenotes}
\end{threeparttable}
\end{table*}

\section{Results}
\subsection{Quality of the fits}

\begin{figure*}
  \centering
 	\includegraphics[width=0.88\textwidth,page=1]{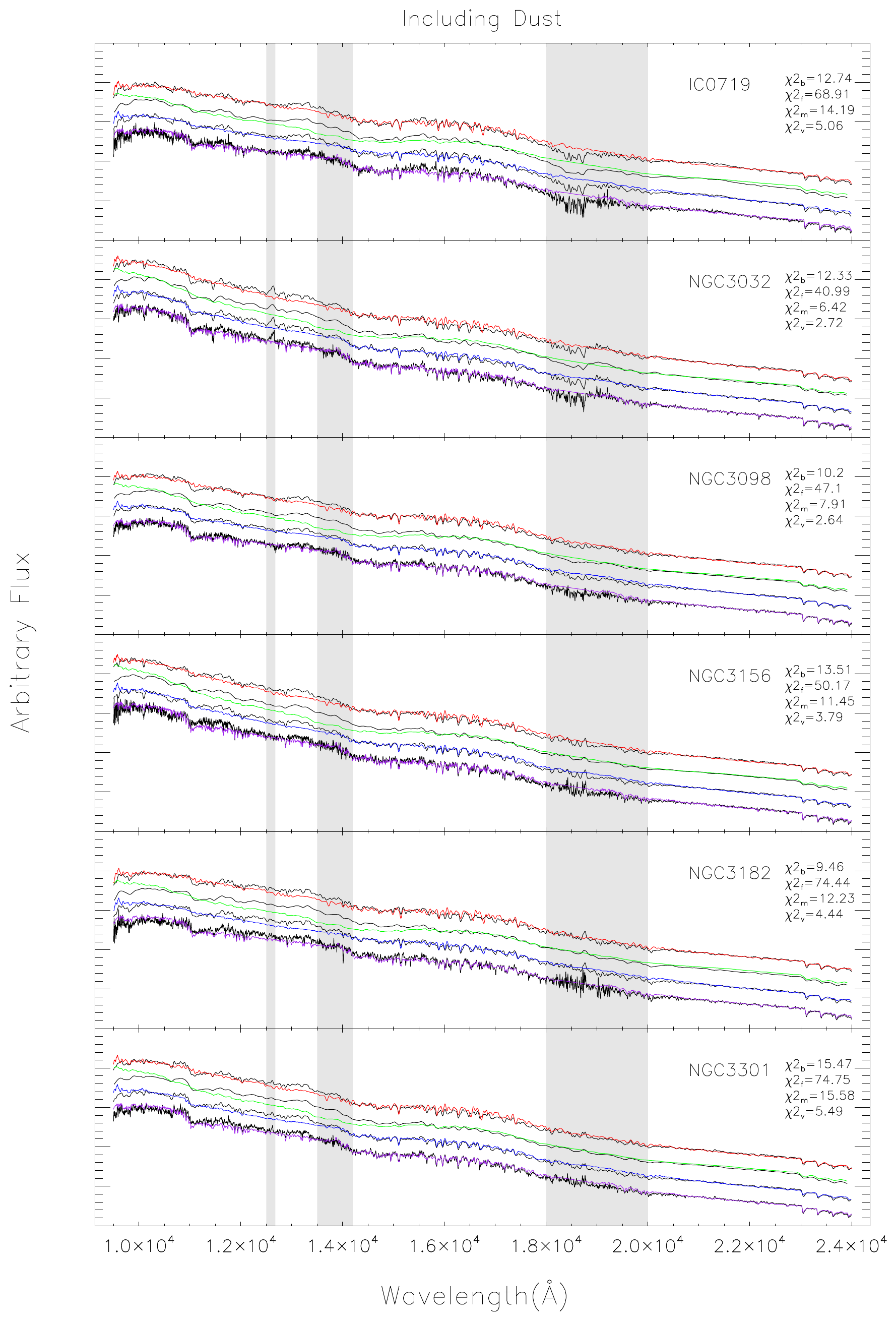}
	\caption{Comparison of spectral fits using the various models for 6 galaxies in the sample. Fits are carried out with a Calzetti extinction curve. The BC03 best fitting template is plotted in red, FSPS in green, M11 in blue and V16 in purple. Regions masked due to strong atmospheric absorption are shown in grey.}
 \label{fig:redfit1}
\end{figure*}

First we focus on how well each model is able to fit the observed near-infrared data. 
\Cref{fig:redfit1}  shows the fits obtained for the first half of the sample using a Calzetti extinction curve, and \Cref{tab:fittablered} gives the reduced $\chi2$ values (\Chi) for these fits, as well as the mass-weighted mean age and the $E(B-V)$ value.

As discussed in section \Cref{sec:noise}, the \Chi\/ values can be used to rank the models in terms of how well they are able to fit the data. \Cref{tab:fittablered} shows that when using an extinction curve the V16 models give the best fits (mean \Chi = $4.9\pm2.8$), followed by M11 and BC03 (mean \Chi = 14.2  and 14.3 respectively), while FSPS fit the worst (mean \Chi = 83.2). 
Inspection of \Cref{fig:redfit1} shows that the V16 models generally fit very well across the entire near-infrared range. M11 and BC03 also do a good job at reproducing the overall shape of the spectra, although there are particular areas where the match is poorer (typically around  1.3~\micron\/ in both models). 
This could be due to the Pickles stellar library, which is input to both models. As previously mentioned, the Pickles library does not have complete spectral coverage for all stellar types over the entire near-infrared wavelength range. 
The FSPS models are the most different to the data in terms of overall spectral shape, which is reflected in the poor values of \Chi\/ obtained with these models. FSPS also consistently requires the largest amount of reddening to match the data. This is inconsistent with SDSS imaging, which shows many of the galaxies have no obvious sign of dust. In many cases, the V16 and BC03 models can match observed shape without invoking any reddening at all, with the notable exception of NGC\,4710, which is visibly dusty at both optical and near-infrared wavelengths. 
This mismatch is attributable to the low resolution BaSeL library used as input to these models, as we show in \Cref{sec:tracklib}.
The fact that BC03, V16 and M11 are able to well reproduce the overall shape of the data indicates that the flux calibration is likely not significantly wrong in either the data or models, or at least is not responsible for all the apparent differences between the models.

\begin{figure*}
  \centering
  	\includegraphics[width=0.9\textwidth,page=1]{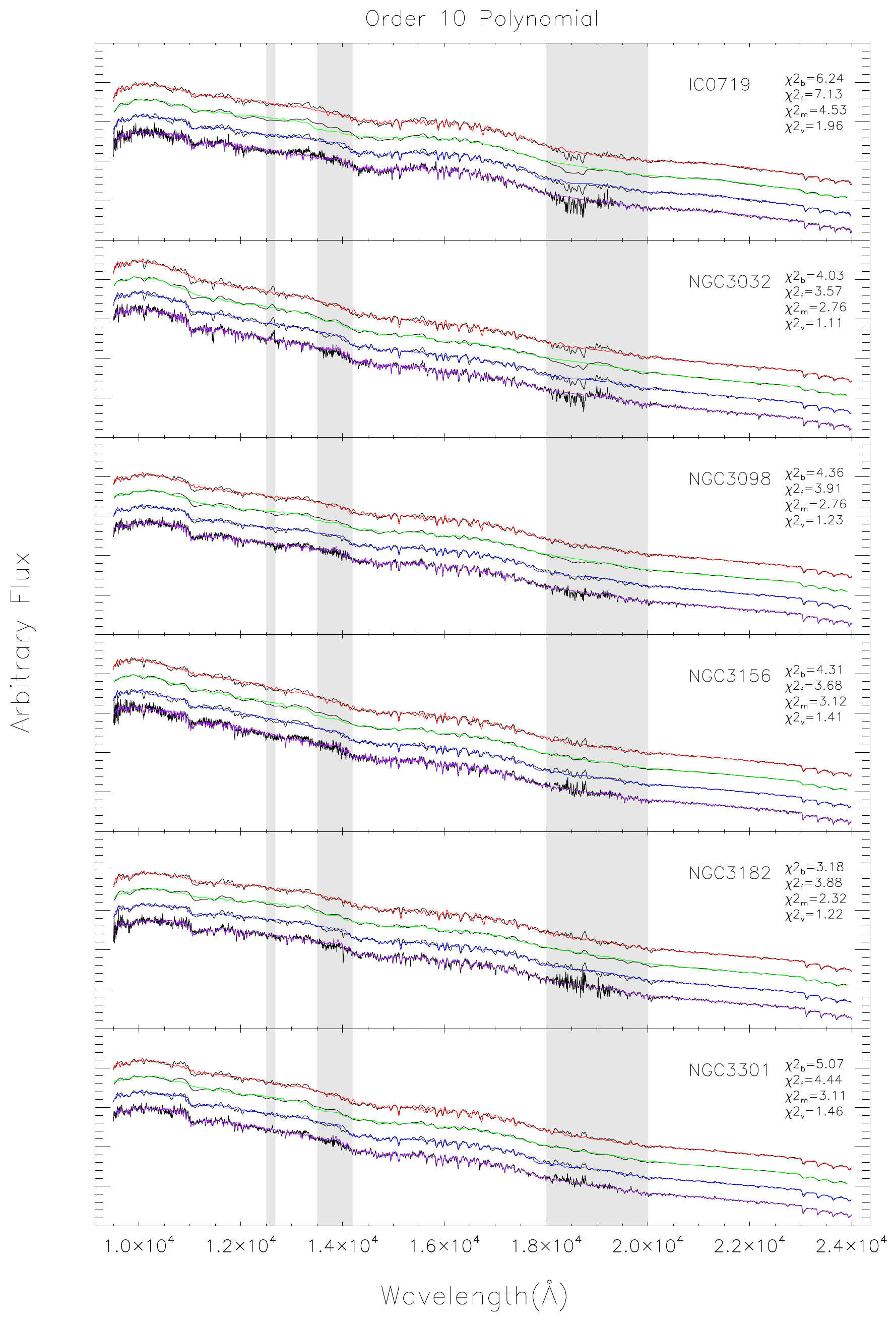}
 	\caption{As in \Cref{fig:redfit1}, but with a multiplicative polynomial instead of extinction curve.}
 \label{fig:mdeg10fit1}
\end{figure*}

\begin{figure*}
  \centering
 	\includegraphics[width=0.9\textwidth,page=1]{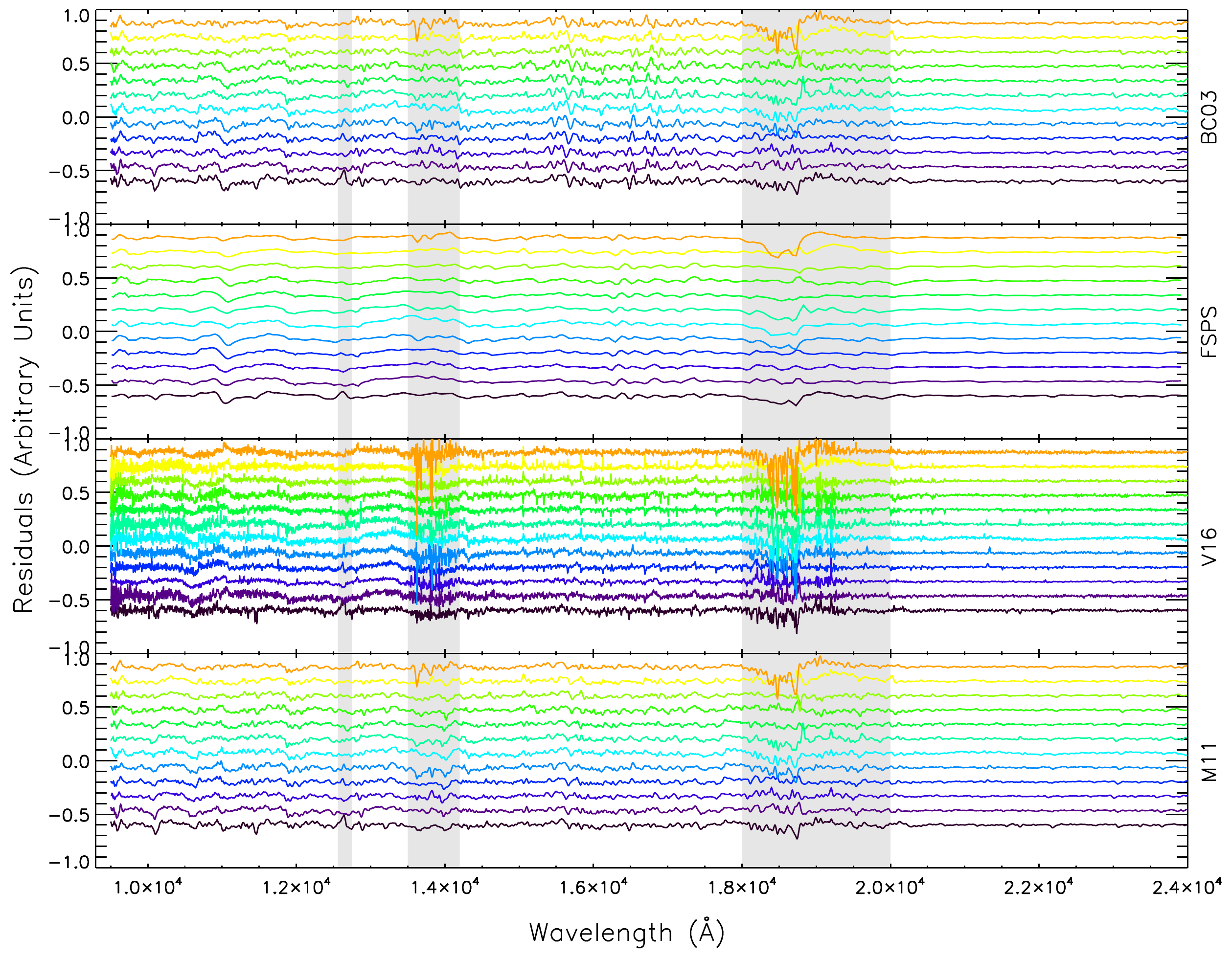}
 	\caption{Fit residuals for each model. Fits were carried out with an order 10 multiplicative polynomial. Each panel shows residuals for a particular model set (top to bottom : BC03, FSPS, V16, M11). Galaxies are ordered by age, with age increasing upwards. Masked regions are shown in grey. Note that the spectral resolution varies strongly between models.}
 \label{fig:mdegresfit}
\end{figure*}

Fitting with a polynomial alleviates the problem of shape mismatch between models and data, and the quality of fit becomes driven by the ability of the models to match the strengths of specific absorption features. 
\Cref{fig:mdeg10fit1} shows the fits obtained for the first half of the sample using a multiplicative polynomial of order 10, and \Cref{tab:fittablemdeg} gives the \Chi\/ values and mean ages.
The ranking of the models in terms of their \Chi\/ values now becomes V16 (mean \Chi $ = 1.5\pm0.3$), followed by M11 (mean \Chi $ = 3.6\pm0.8$), FSPS (mean \Chi $ = 4.5\pm1.0$), BC03 (mean \Chi $= 5.0\pm1.0$). Inspection of \Cref{fig:mdeg10fit1} and \Cref{fig:mdegresfit}, which shows the residuals for each fit, allows us to pinpoint which specific features are poorly fit by the various model sets. 
The BC03 models are unable to  fit the CN feature at 1.1\micron\/, as these models do not contain this feature. The fit to the numerous CO features in H- and K-band, as well as a number of atomic features in K band are also not perfectly matched. FSPS are also unable to fit CN1.1 well, despite containing a CN feature in the models, as the shape of the models in this region does not look similar to the data. V16 fit very well across the entire spectral range, but slightly overpredict CN1.1 strength at all ages. 
M11 also typically fit most features across the spectral range well, slightly underpredicting the strength of CN~1.1\micron\/ and the CO bandheads at 1.6\micron\/ and  2.3\micron.
There is a significant residual visible in all model fits of the two post-starburst galaxies (NGC\,3032 and NGC\,3156) at 1.005~\micron. This is most likely Pa-$\delta$ 
absorption from A-type stars, which have lifetimes of $\sim100$~Myr, and thus are not included in our selected set of models, for which the youngest template is 1~Gyr. This is further evidence that these two galaxies experienced a  recent burst of star formation, 
supporting the strong H$\beta$ absorption visible in the optical spectra of these galaxies shown in \Cref{fig:opt}.

\subsection{Comparison of Star Formation Histories}

\begin{figure*}
  \centering
 	\includegraphics[width=0.9\textwidth,page=1]{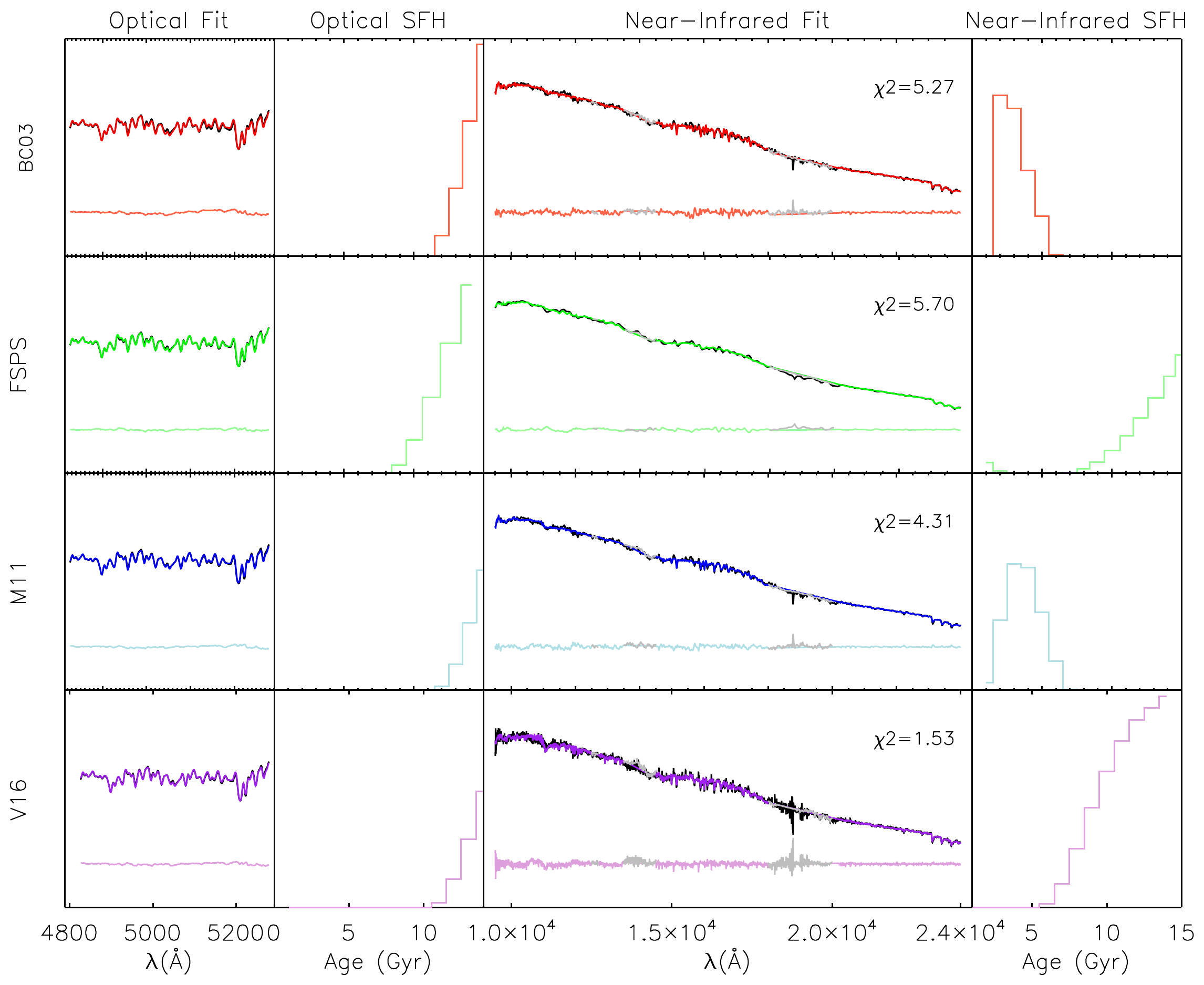}
 	\caption{NGC\,4578: Comparison of spectral fits and star formation histories  obtained from the four models, as labelled on the left axis, using a multiplicative polynomial. Panels show (from left to right) the optical spectra: observed (black), best fit and residuals (coloured); the star formation history derived from the optical fit; the near-infrared spectra (coloured as before); the star formation history derived from the near-infrared fit. }
 \label{fig:sfhfit}
\end{figure*}

\begin{figure}
  \centering
  	\includegraphics[width=\columnwidth]{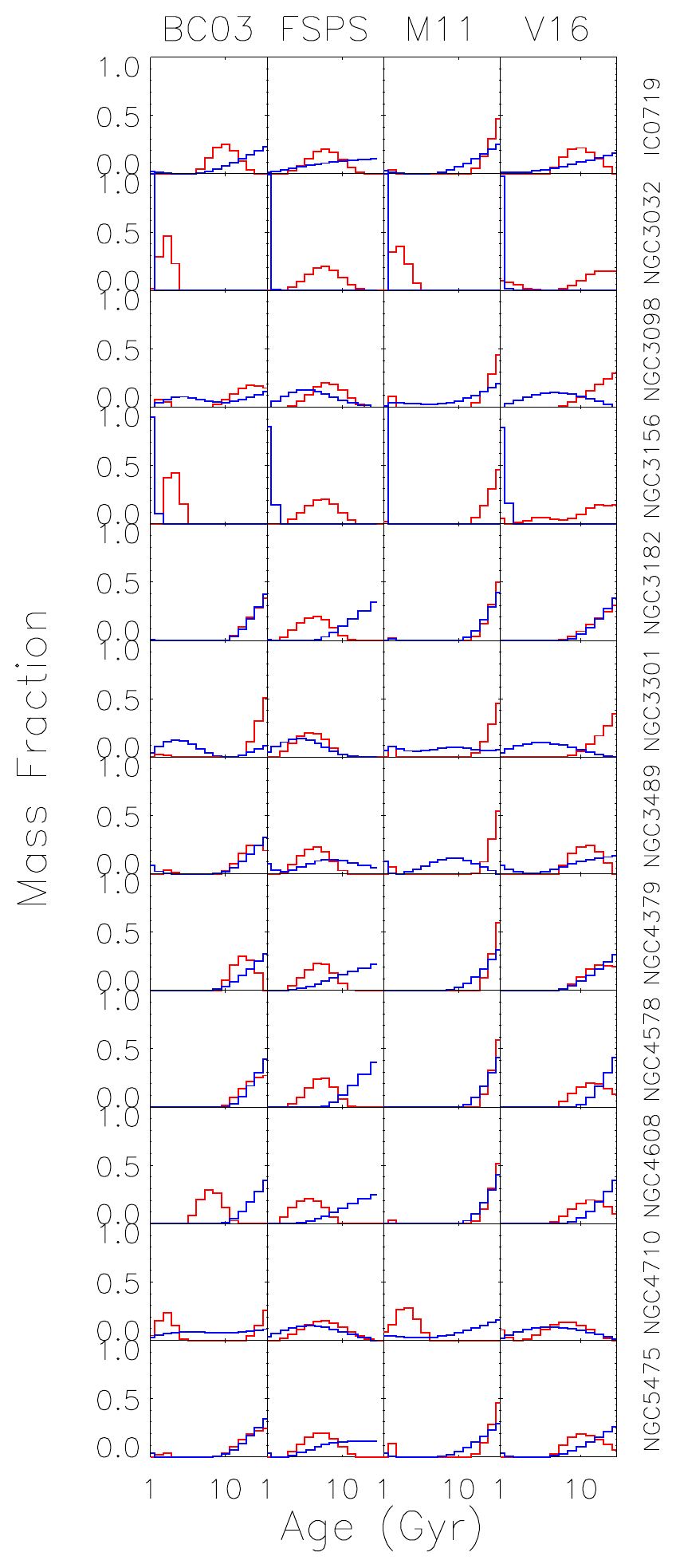}
 	\caption{Star formation histories derived using a Calzetti extinction curve. Optically derived SFHs are plotted in blue, and near-infrared derived SFHs in red. }
 \label{fig:redsfh}
\end{figure}

\begin{figure}
  \centering
   	\includegraphics[width=\columnwidth]{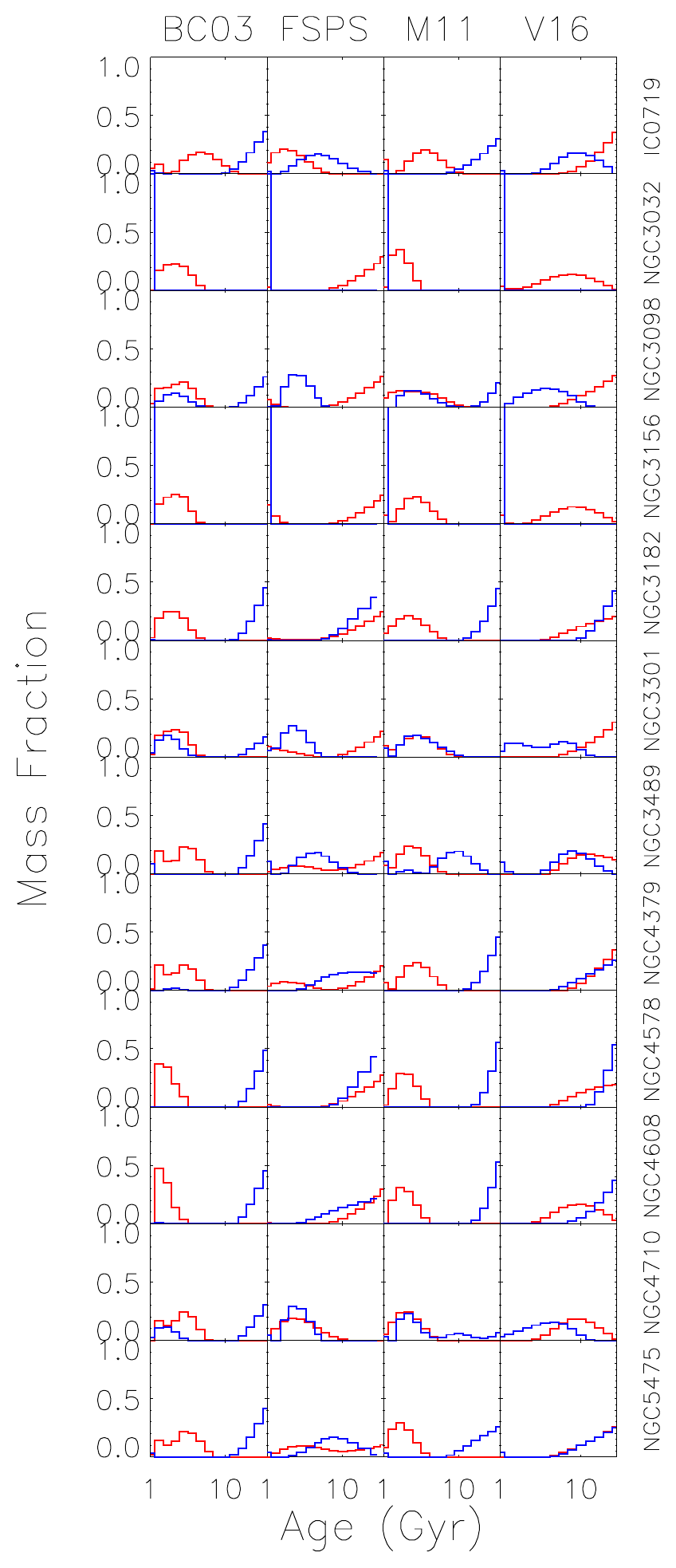}
 	\caption{As in \Cref{fig:redsfh}, but with a multiplicative polynomial in place of extinction curve.}
 \label{fig:mdeg10sfh}
\end{figure}

We next compare the galaxy ages and star formation histories that are derived using the various models. 
\Cref{fig:sfhfit} shows representative optical and near-infrared fits, and the SFHs derived from them, for NGC\,4578, using a multiplicative polynomial.
The optical fits typically have lower \Chi\/ than the near-infrared fits. SFHs derived from optical spectroscopy of all models typically agree very well, and correlate well with the SSP ages from \atlas\/ \citep{mcdermid_atlas3d_2015}. In particular, all models derive young ages for our two post-starburst galaxies (NGC\,3032 and NGC\,3156) when fitting their optical spectra. 
When fitting near-infrared spectra alone, however, the derived SFHs are model dependent, and also depend strongly on treatment of the continuum. 
\Cref{fig:redsfh,fig:mdeg10sfh} show the SFHs derived for the entire sample from both wavelength ranges. The optical SFHs are plotted in blue and the near-infrared SFHs in red. As the results depend on the treatment of the continuum, we analyse the two cases separately.

\subsubsection{Reddening case}

\begin{figure*}
  \centering
  	\includegraphics[width=\textwidth]{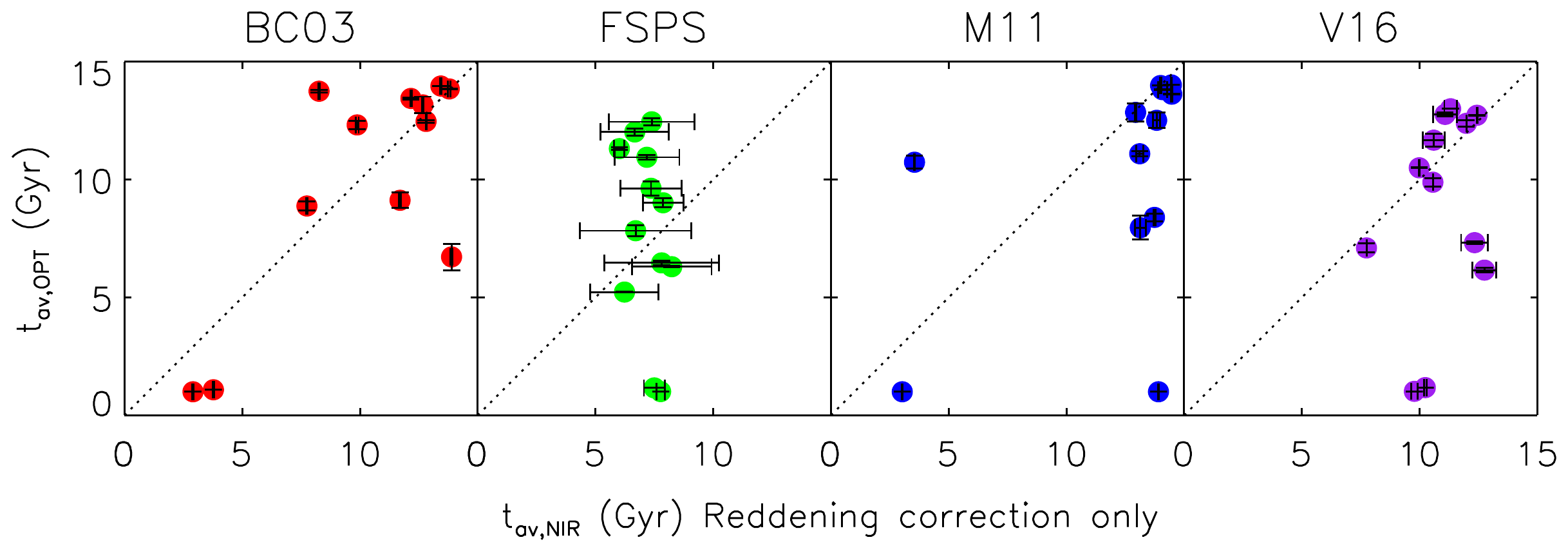}
  	\includegraphics[width=\textwidth]{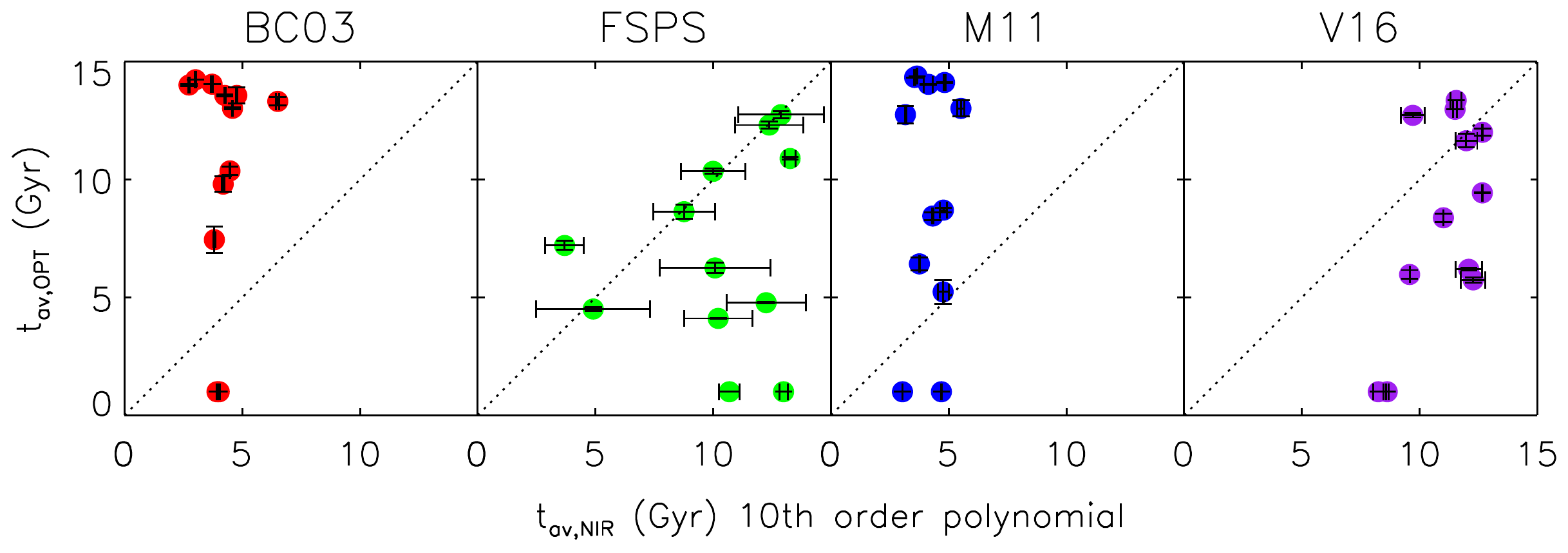}
 	\caption{Comparison of mass-weighted mean ages derived using the optical and near-infrared spectra for each model set. Fits are carried out with a Calzetti extinction curve (top) and multiplicative polynomial (bottom). Note that the spectral library used in FSPS here is BaSel, compared with \Cref{fig:tav_fsps_red}, which uses an updated spectral library.}
 \label{fig:meanage}
\end{figure*}

Regardless of wavelength range, BC03 identifies post-starbursts  NGC\,3032 and NGC\,3156 as young, and NGC\,4710 as containing a substantial fraction of young stars. Both wavelengths also require smaller fractions of young stars in NGC\,3098, NGC\,3301, NGC\,3489 and NGC\,5475. 
The remainder of the galaxies are solely old in both wavelength ranges, with the curious exception of NGC\,4608, which appears younger in the near-infrared than it does in the optical.
Use of the M11 models in the near-infrared also leads to young ages for NGC\,3032 and NGC\,4710, in agreement with the optical, however the post-starburst galaxy NGC\,3156 appears old in this case. 
As with BC03, M11 models require small fractions of young stars in NGC\,3098, NGC\,3301, NGC\,3489, NGC\,4608 and NGC\,5475. The remainder of the galaxies appear old at both wavelengths. 
The V16 near-infrared fits imply a small burst in NGC\,3032, atop a pre-existing old population, which is the most physically likely scenario. 
NGC\,3156 and NGC\,4710 also have extended SFHs, however the mass-weighted mean ages derived using V16 are all old ($t\gtrsim$~8yr), contrary to the optical SFHs, which span a large range of mean ages. Galaxies such as NGC\,3098 and NGC\,3301 have rather inconsistent SFHs when comparing optical and near-infrared results.
The near-infrared FSPS fits do not appear to be very strongly age sensitive, with most galaxies appearing `middle-aged' ($t\approx6-8$~Gyr), regardless of the optical ages (but see \Cref{sec:tracklib}, where we show that this is likely due to the BaSeL library used in these models).
In \Cref{fig:meanage}, we summarise these results by comparing, for each model, the mass-weighted ages of our sample derived from both wavelength ranges.

\subsubsection{Polynomial case}

While use of a reddening law with the M11 and BC03 models led to SFHs  which bore some resemblance to the optical ones, modelling the continuum with a multiplicative polynomial causes these models to effectively lose all sensitivity to age in the near-infrared. 
In this case, the BC03 and M11 SFHs all look incredibly similar, giving 
mass-weighted mean ages which all lie between $\approx3-7$~Gyr
despite the sample spanning $\approx13$~Gyr in optically derived mass-weighted age (see \Cref{fig:meanage}). 
The V16 near-infrared SFHs are similar to those derived with reddening:  all galaxies appear largely old, but NGC\,3032 and NGC\,3156 have somewhat more extended SFHs, resulting in mean ages of  $\approx8$~Gyr for these two, which, while still old, are the youngest in the sample. The FSPS SFHs display the opposite behaviour to the other three models: in this case the polynomial fits result in SFHs that vary strongly from galaxy to galaxy, however they do not display a clear relationship with the optically derived SFHs.

\subsection{Model Ingredients}
This section explores the effects of various model ingredients on derived physical properties by carrying out spectral fits using different model permutations. From now on, all fits are done with a Calzetti extinction curve rather than a polynomial. 

\subsubsection{Non-solar metallicities }

M11 and BC03 predictions using the Pickles library are available only for solar metallicity. So as to compare the various model sets on an equal basis, we fit using only solar metallicity templates for every model. 
The sample was selected (based on optical SSP-equivalent parameters) to be approximately solar metallicity, so we expect this basis set to be adequate. We test this assumption by extending the metallicity coverage using the V16 models, which are available for multiple metallicities.

\begin{figure}
  \centering
  	\includegraphics[width=\columnwidth]{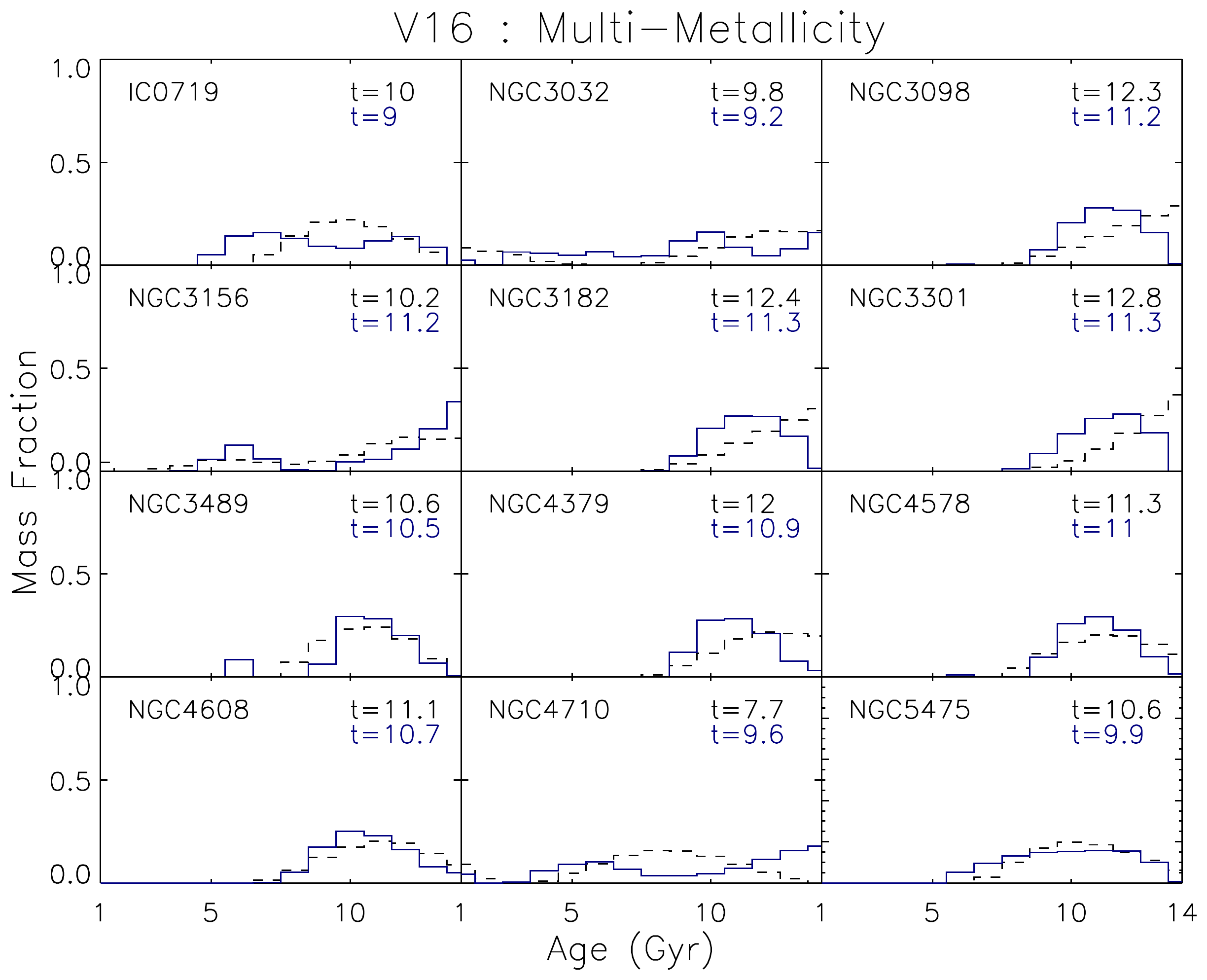}
 	\caption{Effect of extending metallicity coverage using V16 models. The SFHs derived using our default basis set (solar metallicity only) are shown as black dashed lines, and the SFHs derived allowing metallicity to vary at each age from -0.35 to 0.26 are shown in blue. The derived mass-weighted mean ages are printed in each panel in Gyr. SFHs are derived using a reddening law.}
 \label{fig:multiz}
\end{figure}

\Cref{fig:multiz} shows the SFHs derived using the default V16 basis set (black, solar metallicity only) and a basis set including five different metallicities from $-0.35\leq [{\rm Z}/\zsun]\leq 0.26 $ (blue). For most galaxies, the difference is minimal.  
The difference in mean ages is typically less than 1~Gyr, with the addition of multiple metallicities usually leading to a marginally younger mean age. The SFHs themselves are very similar. The largest differences are seen in IC\,0719 NGC\,3032, NGC\,3156 and NGC\,4710, which look burstier when fitting with multiple metallicities. These galaxies all have clear signs of recent star formation at other wavelengths. As such, they are expected to contain young, super-solar populations. Even so, the results obtained using the extended basis set are not substantially different to those obtained using solar metallicity only. 

\subsubsection{Spectral libraries and stellar tracks}
\label{sec:tracklib}
Most current model sets are available for a variety of input stellar tracks and spectral libraries which the user can choose from. Here we include a brief analysis of the effect of these choices on the derived stellar population properties. 

\begin{figure}
  \centering
  	\includegraphics[width=\columnwidth]{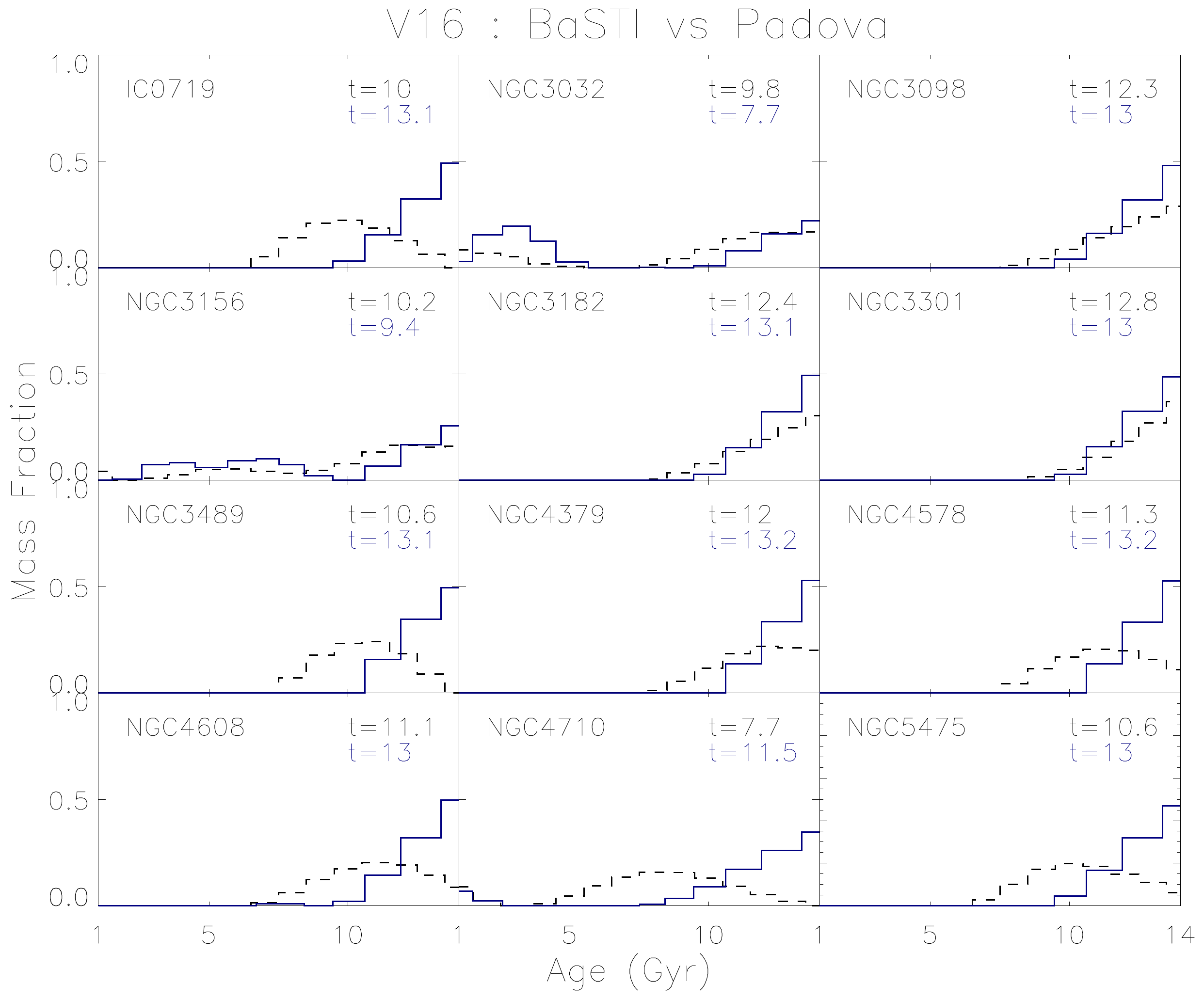}
 	\caption{Effect of varying tracks using V16 models. The SFHs derived using our default tracks (BaSTI) are shown as black dashed lines, and the SFHs derived using the Padova tracks are shown in blue. The mass-weighted mean ages are printed in each panel in Gyr. SFHs are derived using a reddening law. }
 \label{fig:tracks}
\end{figure}

First, we study the impact of stellar tracks. \Cref{fig:tracks} shows the SFHs derived with the V16 models using our default choice of the BaSTI tracks (black), as well as when using the Padova tracks (blue). 
There is a small but noticeable change in the derived SFHs, with the Padova tracks resulting in more extreme SFHs i.e. when using the Padova tracks, the post-starburst galaxies appear younger while the rest of the sample appear older. The change in the mass-weighted mean ages is typically $\approx2$~Gyr. The older ages are due to a known effect in the Padova (2000) tracks, whereby a hotter RGB at old ages leads to 
a bluer colour in these tracks compared to previous works by the Padova group, as well as other popular stellar evolution calculations. This means that for the same colour, use of the Padova (2000) tracks will result in an older age (see e.g. \citet{bruzual_stellar_2003}).

\begin{figure}
  \centering
  	\includegraphics[width=\columnwidth]{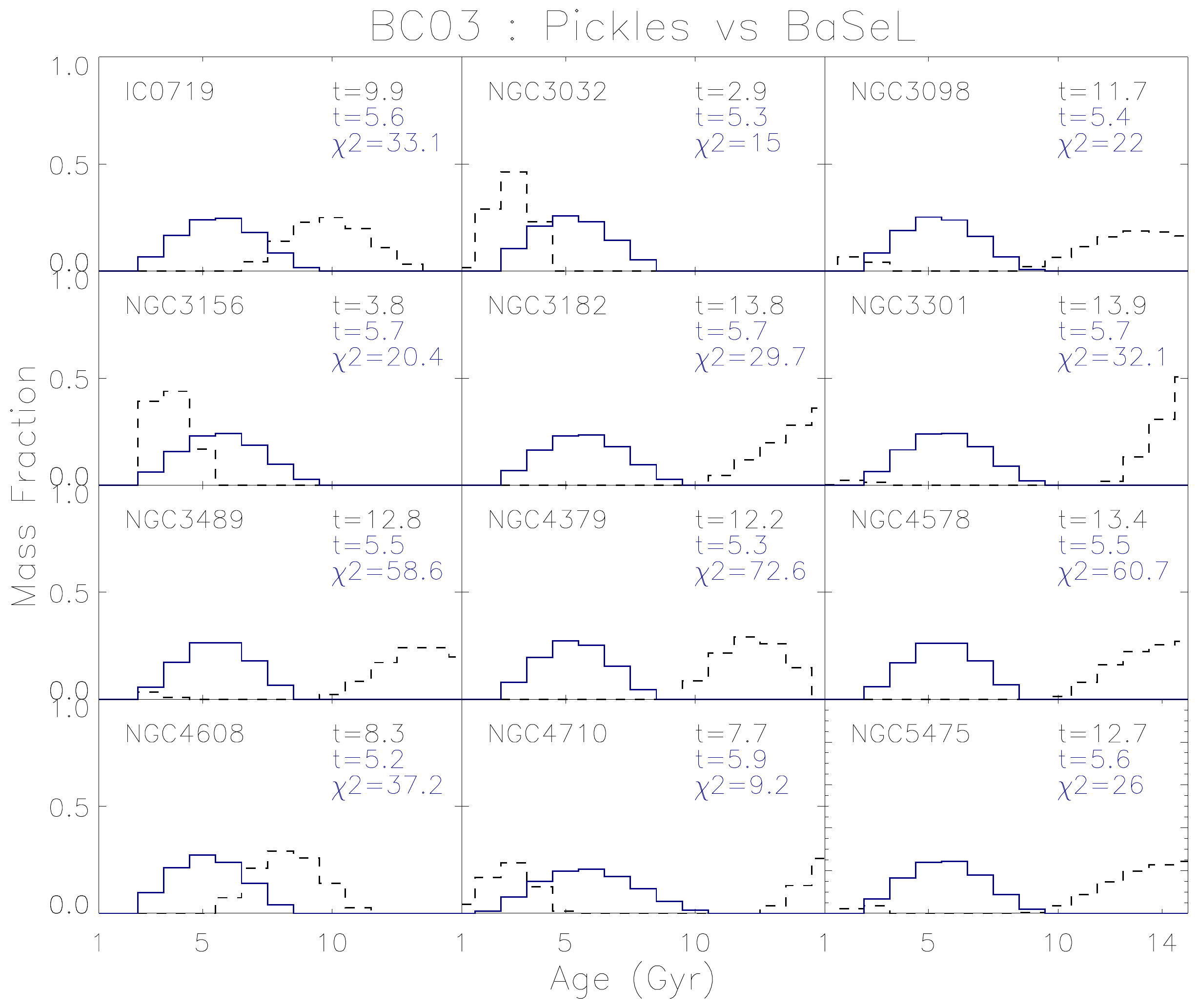}
 	\caption{Effect of varying spectral libraries using BC03 models. The SFHs derived using our default library (Pickles) are shown as black dashed lines, and the SFHs derived using the BaSeL library are shown in blue. SFHs are derived using a reddening law. The \Chi\/ of the BaSeL fits are printed in each panel, as well as the mass-weighted mean ages calculated for each library.}
 \label{fig:libs}
\end{figure}

Turning next to the choice of spectral libraries: we originally used BC03 with the Pickles empirical library (the same library used by M11), however the default BC03 library at near-infrared wavelengths is the Basel theoretical library (the same library used by the default FSPS). The FSPS fits had the largest \Chi\/ and the resultant SFHs displayed the lowest correlation with the optical SFHs. 
To test whether this was due to choice of spectral library, we carried out full spectral fitting using BC03 with BaSeL. 
\Cref{fig:libs} shows the SFHs derived with an extinction law using BC03-Pickles (black) and BC03-BaSeL (blue). There is a significant discrepancy between the two. When using BC03 with the BaSeL library, 
the derived SFHs are almost identical to those found previously using FSPS.
The \Chi\/ (plotted in each panel) are high (\Chi $= 34.7\pm20$) and the  SFHs are essentially insensitive to age, indicating that the BaSeL library was likely the primary cause of the of the poor behaviour seen when using the FSPS models. 

\begin{figure*}
  \centering
  	\includegraphics[width=\textwidth]{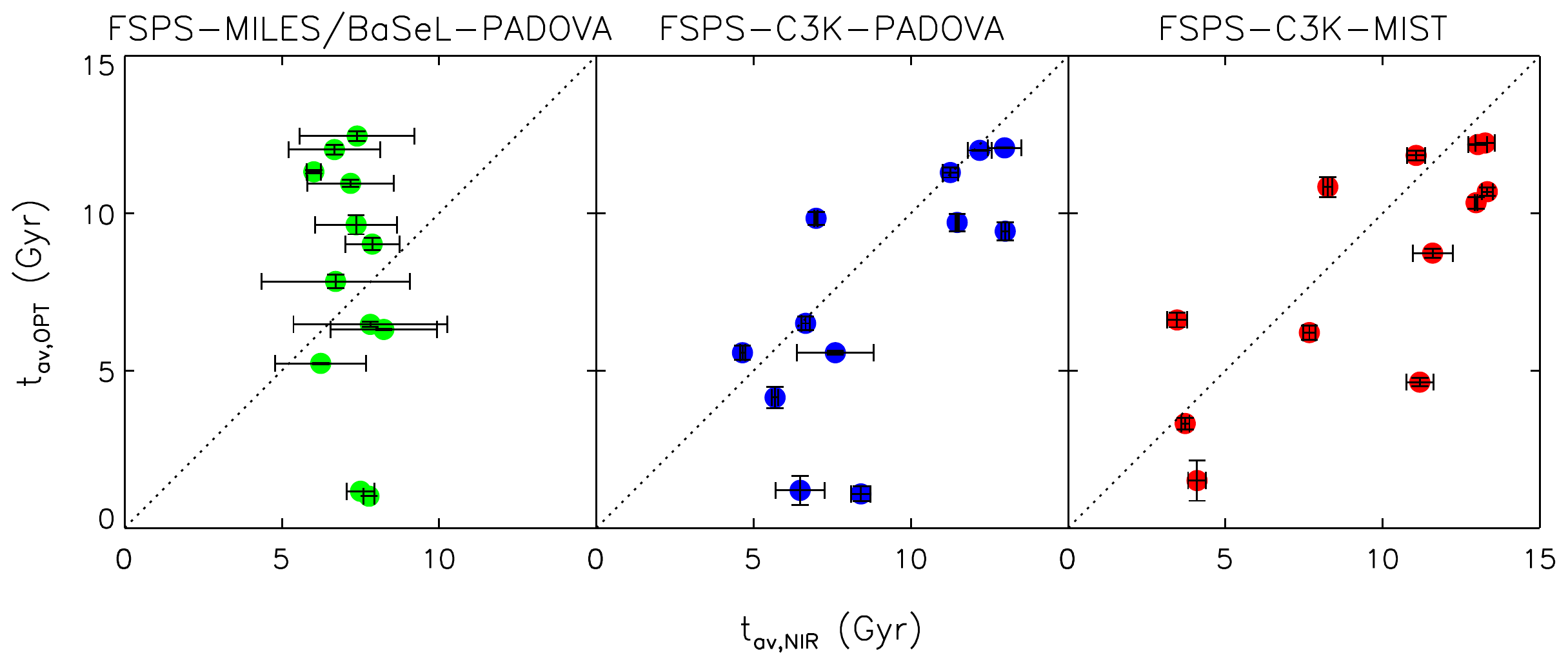}
 	\caption{Comparison of mass-weighted mean ages derived from optical and near-infrared fitting with various FSPS setups, using a Calzetti extinction curve.}
 \label{fig:tav_fsps_red}
\end{figure*}

\begin{figure*}
  \centering
  	\includegraphics[width=\textwidth]{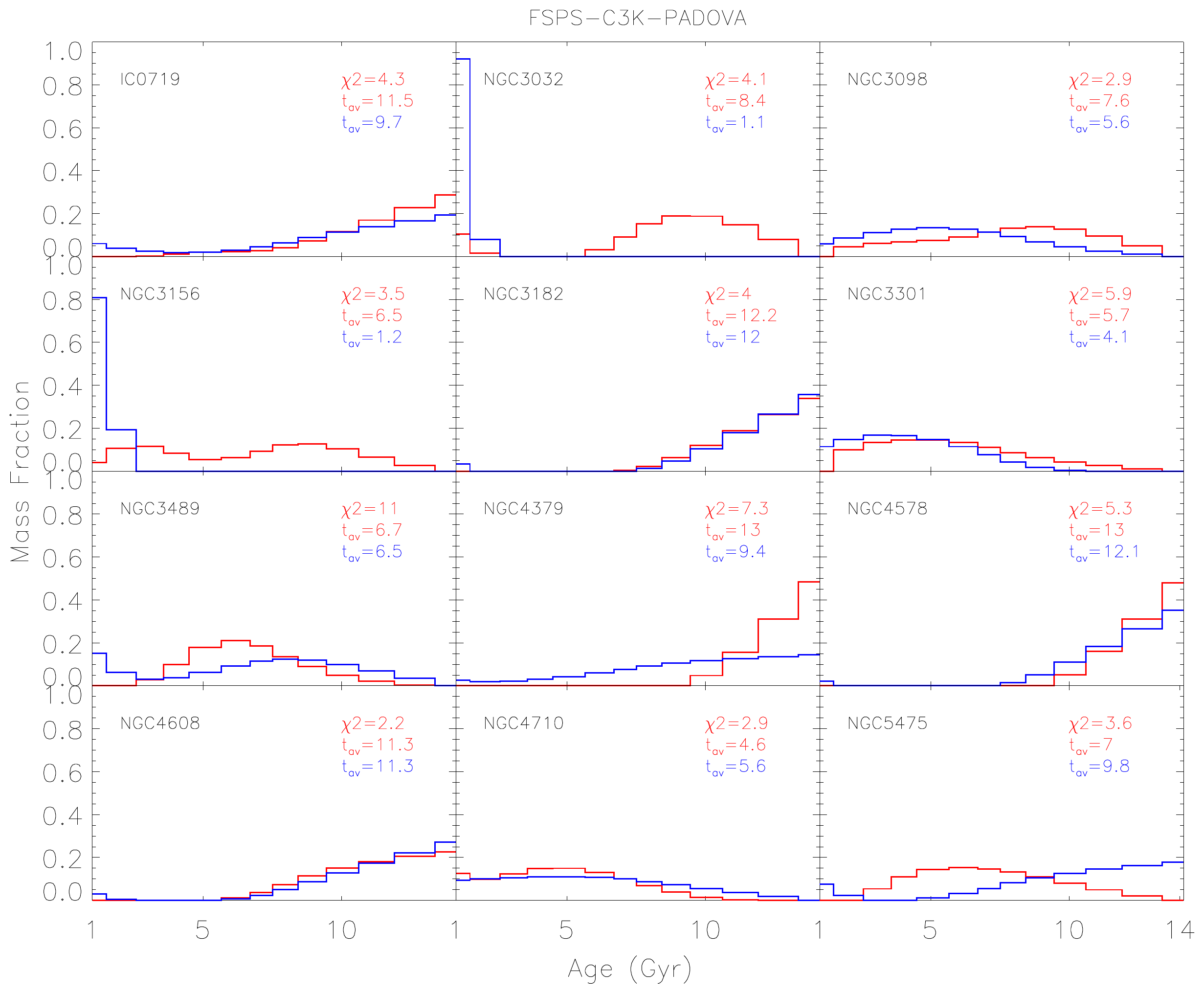}
 	\caption{SFHs derived from optical (blue) and near-infrared (red) spectroscopy using FSPS-C3K-PADOVA models. The mass-weighted mean age derived in the near-infrared (optical) is given in each panel in red (blue) and the \Chi\/ of the near-infrared fit is printed in red.  }
 \label{fig:fspssfh}
\end{figure*}

\begin{figure*}
  \centering
  	\includegraphics[width=\textwidth]{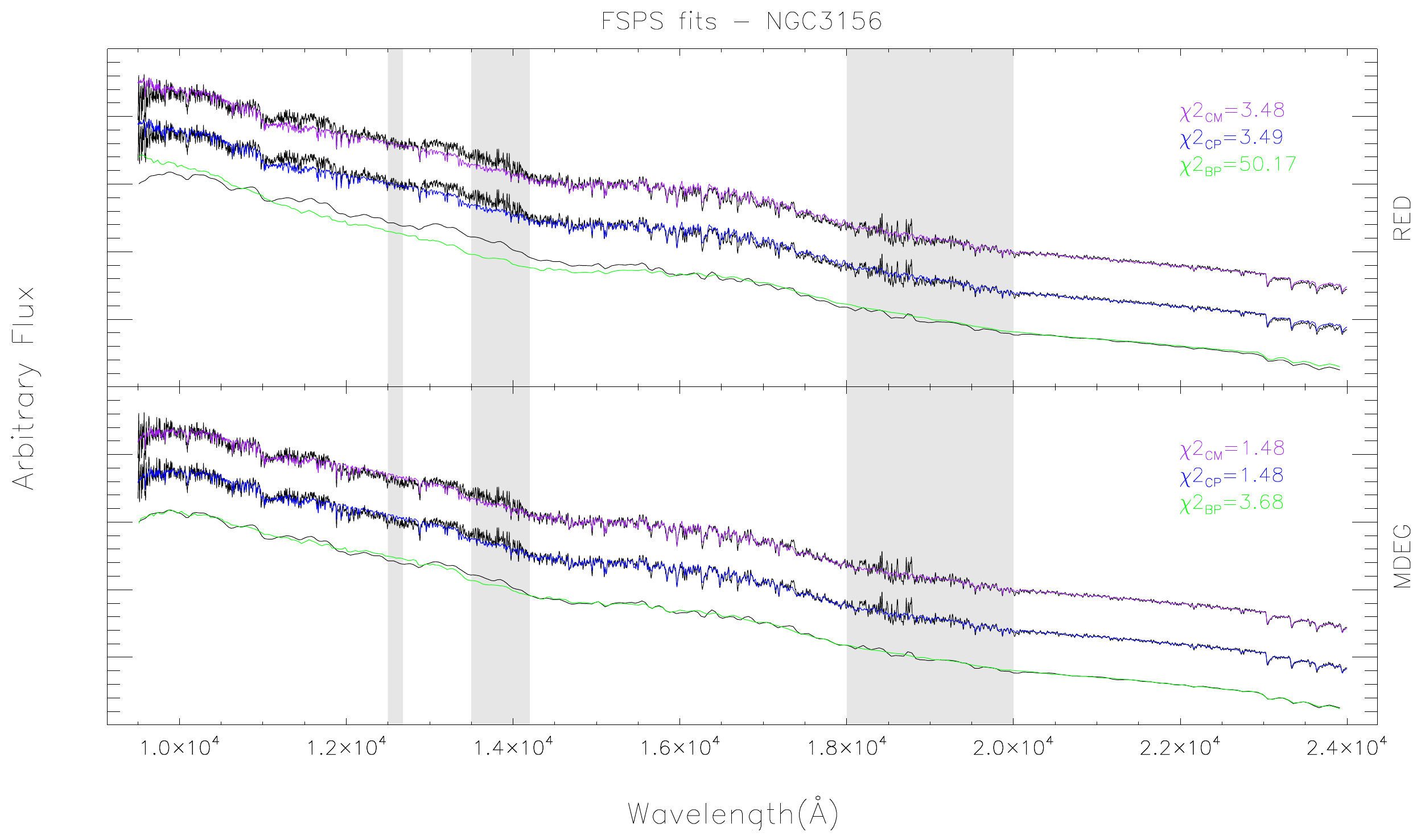}
 	\caption{Comparison of spectral fits using the various FSPS versions. Fits are carried out with a Calzetti extinction curve. From top to bottom: FSPS-C3K-MIST (update; purple), FSPS-C3K-PADOVA (update; blue), FSPS-BASEL-PADOVA (default; green). Masked regions are shown in grey. }
 \label{fig:fspsfits}
\end{figure*}

The FSPS models have recently been updated \citep[see][]{hutchinson_redshift_2016} to include a new, high-resolution theoretical spectral library (C3K) as well as new MESA Isochrones and Stellar Tracks \citep[MIST;][]{choi_mesa_2016}, and as a final test, we fit the sample using both of these updates. \Cref{fig:tav_fsps_red} shows the mean ages derived from both optical and near-infrared spectroscopy using various FSPS setups with an extinction law. While the current publicly available FSPS models have no age sensitivity in the near-infrared, the situation drastically improves with use of the updated high resolution C3K library. Using FSPS-C3K-Padova on both wavelength ranges, one obtains better agreement between mass-weighted mean ages for the oldest galaxies. Use of FSPS-C3K-MIST brings the youngest galaxies into better agreement as well. 
The SFHs derived using FSPS-C3K-Padova are shown in \Cref{fig:fspssfh}. The agreement between optical and near-infrared SFHs is much better than seen previously. The largest discrepancies are seen in the post-starburst galaxies. Optically, these galaxies were formed entirely in the last 2~Gyr. The near-infrared SFHs contain young bursts on top of a substantial old component. This is a more physically realistic scenario than that implied by the optical - the optical result is likely due to the old population being outshone by the young stars at these wavelengths. 
The updated FSPS-C3K models give \Chi\/ values  comparable to V16 (mean \Chi $ = 4.8\pm2.4$ with an extinction curve). An example fit is shown in \Cref{fig:fspsfits} for each of the model versions and the derived properties are given \Cref{tab:fspsproperties}.

To summarise: the largest effect on the derived SFHs comes from the choice of stellar spectral library. Choice of stellar tracks, and metallicity coverage have only secondary effects on the SFHs derived from our near-infrared spectroscopy.

\subsection{Line strengths}
\label{sec:ls}

\begin{figure*}
  \centering
  	\includegraphics[width=\textwidth]{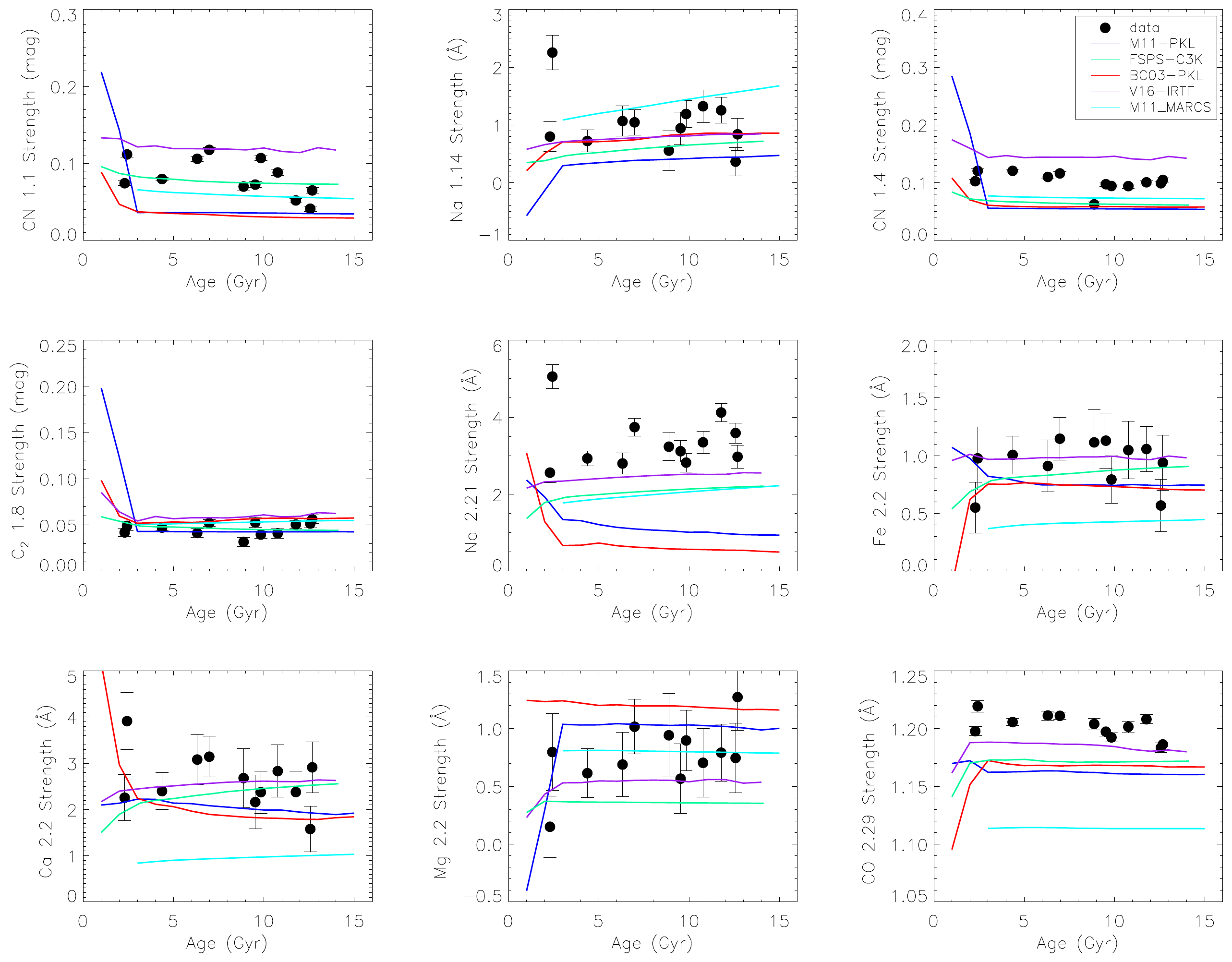}
 	\caption{Strengths of chosen indices measured in the data, compared with the predictions of SPS models. The model predictions are plotted with coloured lines as a function of SSP age. The sample measurements are plotted against the galaxies' optically-derived mass-weighted ages.}
 \label{fig:ls}
\end{figure*}

\Cref{fig:ls} shows the strengths of our chosen indices compared with the predictions of the four models. Model predictions are plotted as coloured lines, and the galaxies (filled circles) are plotted 
against their optically-derived mass-weighted ages from \atlas\/ 
\citep{mcdermid_atlas3d_2015}. 
These mass-weighted ages are less dominated by young stars, and thus more representative of the  ages of the dominant population in the galaxies. 
All indices were measured at the resolution of the Pickles library (BC03/M11). The FSPS predictions plotted here as those using the C3K library. We chose to use this rather than the default FSPS-BaSeL, as those models are not line-by-line radiative transfer models, and are thus not suitable for detailed spectral predictions. 
As mentioned earlier, the Pickles library used in M11 and BC03 lacks spectroscopic observations for many of its stars, and as such, 
some near-infrared absorption features may be poorly resolved. We still include the measured Pickles line strengths, but for comparison, we also include index strengths measured in 
the M11 model in their version based on the MARCS library. This comprises a comprehensive grid of cool stars (2500-8000~K) at very high resolution (R=20,000) over a large wavelength range (1300-20,000~{\AA}), allowing the calculations of SSP models with ages larger than 3 Gyr, for various chemical compositions (see www.maraston.eu/M11). As already noted, creation of model atmospheres in the near-infrared is not a trivial task, due to the multiplicity of lines and molecules to be taken into account when the temperature is low. This comparison can give us a qualitative idea on which indices allow a fair comparison with data. 

\Cref{fig:ls} shows that molecular indices CN1.1, CN1.4, and C$_2$ are predicted by the M11 models to be strong in intermediate age populations ($\leq2$~Gyr) which are affected by the TP-AGB phase. Other models predict no such upturn (or a much milder upturn) at these ages.
The data do not show a strong increase in these indices at the youngest ages, rather the measured index strengths are approximately constant with age (though recall the data show mass-weighted average ages of integrated populations, not true SSPs).
The strength of CN1.4 measured in the data is less than that predicted by V16 at all ages, and greater than predicted by all other model sets at most ages. 
CN1.1 displays similar behaviour, but the strength is accurately modelled by FSPS-C3K. The underprediction of CN by most model sets at old ages is likely due to deficiencies in the spectral libraries used. Use of the M11 models with the theoretical MARCS library rather than Pickles predicts CN1.1 strengths which are in better agreement with FSPS-C3K and the data. 
The strength of C$_2$ in the data is consistent with all models for old ages, despite a range in mass-weighted and SSP-equivalent ages that spans 1-15~Gyr i.e. there is no evidence of a significant contribution from TP-AGB stars in any of the galaxies in the sample. 
The molecular CO strength measured in the MARCS library stands out as being systematically lower than the data. Our CO index definition measures the strength of the first CO feature in the bandhead. 
In the MARCS models, this first CO feature is visibly weaker than the following features, contrary to the behaviour of other models.

Most atomic features are not predicted by the models to vary strongly as a function of age in the near-infrared range. Measurements support this lack of age variation, however the strengths of particular features are not fit equally well by all models. 
The BC03 and M11 models based on the Pickles library are outliers in the measured Na2.21 strength compared to other models. 
Both sodium features (Na1.14 and Na2.21) are somewhat enhanced in the data compared to model predictions (although approximately half the sample is consistent with V16 for Na1.14). 
An enhanced sodium contribution compared to model predictions has been seen in a number of recent works \citep[e.g.][]{smith_imf-sensitive_2015}. Possible explanations for this discrepancy include enhanced [Na/Fe] or a bottom heavy IMF (i.e. an IMF containing more dwarf stars than a canonical Kroupa IMF). 
Multiple studies have found evidence for a bottom heavy IMF in the centres of massive ellipticals \citep[e.g.][and others]{conroy_stellar_2012,ferreras_systematic_2013,spiniello_stellar_2014}, however this is not expected in this sample based on the low $\sigma_e$ of these galaxies. 
NGC\,3032 has unusually high Na and Ca compared to all other galaxies in the sample, which could be due to populations $<1$~Gyr present in the galaxy, although the same enhancement is not seen in NGC\,3156. 

\section{Discussion}

In the sections above, we compared the SFHs derived from optical and near-infrared spectroscopy using various models, as well as comparing the strengths of various near-infrared indices with the predictions of these models. Here we discuss some of the questions raised by this analysis. 

\subsection{Should optical and near-infrared properties agree?}
One of the checks we carried out was to test the various models for self-consistency, which we did by deriving SFHs from both optical and near-infrared wavelength ranges using the same model set.
However, the question may be asked, should the properties derived using different wavelength ranges actually agree with one another?

In theory, if the models are correct, they should give consistent results at all wavelengths. In practice, however, this is not necessarily the case. 
Optical SSP-equivalent properties are known to be heavily biased towards the age of the youngest population present, as young stars are more luminous and blue, thus contributing more to the light-weighted values \citep{trager_stellar_2000-2,serra_interpretation_2007}. The near-infrared, on the other hands, is dominated by stars on the red and asymptotic giant branches, so SSP-equivalent ages derived in the two wavelength regimes may not necessarily agree in galaxies  containing young stars. 

The problem is somewhat alleviated, however, by the use of full spectral fitting to derive non-parametric SFHs. Fitting a linear combination of SSPs (compared to finding the single model template that most closely matches the strengths of particular absorption features) allows one to also include old populations which do not contribute substantially to the light. The situation is improved even further by enforcing regularisation in the fits. Regularisation effectively imposes a penalty on the $\chi2$ when the derived SFH is not smooth. 
While the requisite young populations will still be included in the SFH, the smoothness criterion ensures that the SFH includes as large a fraction of old templates as possible without significantly degrading the fit.
In practice, the optical is still somewhat biased to young ages, 
as the old fraction is difficult to constrain given that the majority of the optical information is provided by the young populations. Inspection of \Cref{fig:redsfh} shows that within a model set, the agreement between the optical and near-infrared SFHs is much worse than the agreement between the optical SFHs of different models. This may result from the different SFH-constraining power of the two spectral regions, at least with the quality of data and models used here.

\subsection{Why does the near-infrared give such inconsistent results?}
\begin{figure}
  \centering
  	\includegraphics[width=\columnwidth]{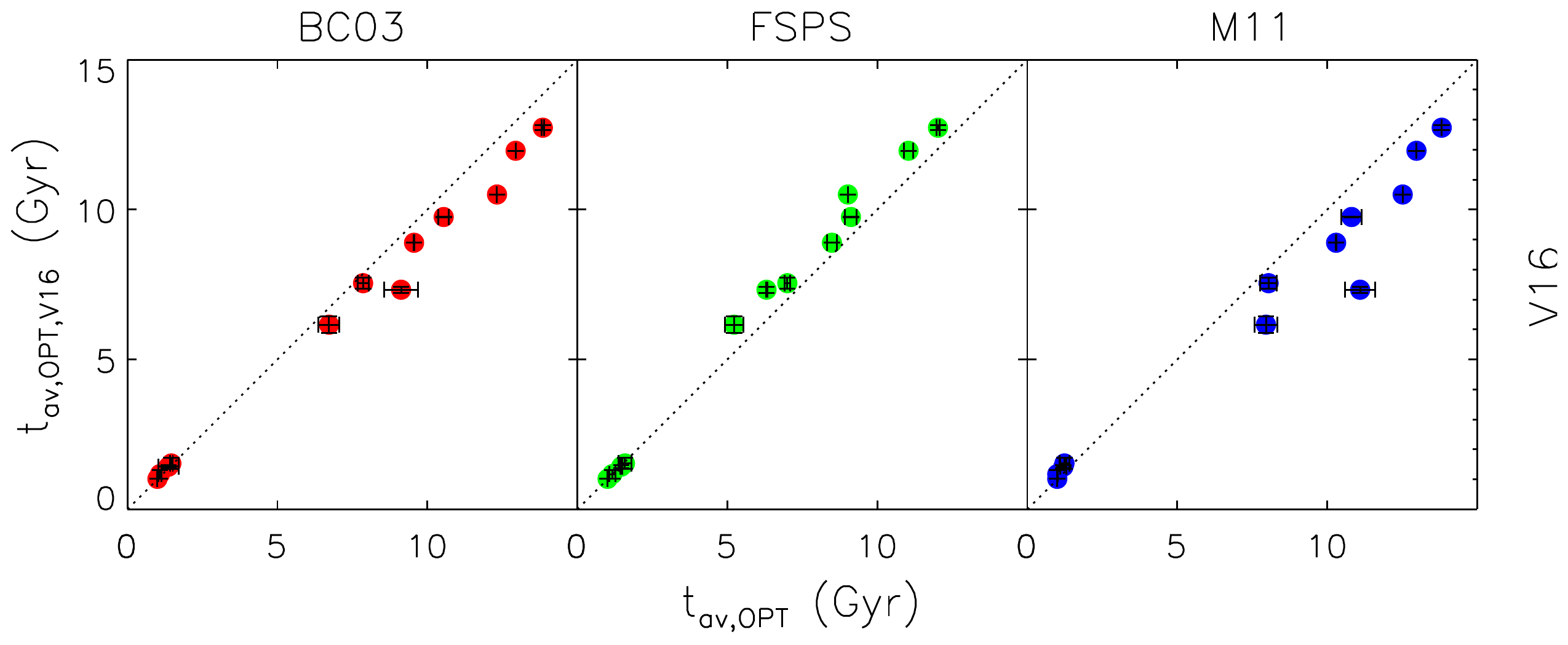}
  	\includegraphics[width=\columnwidth]{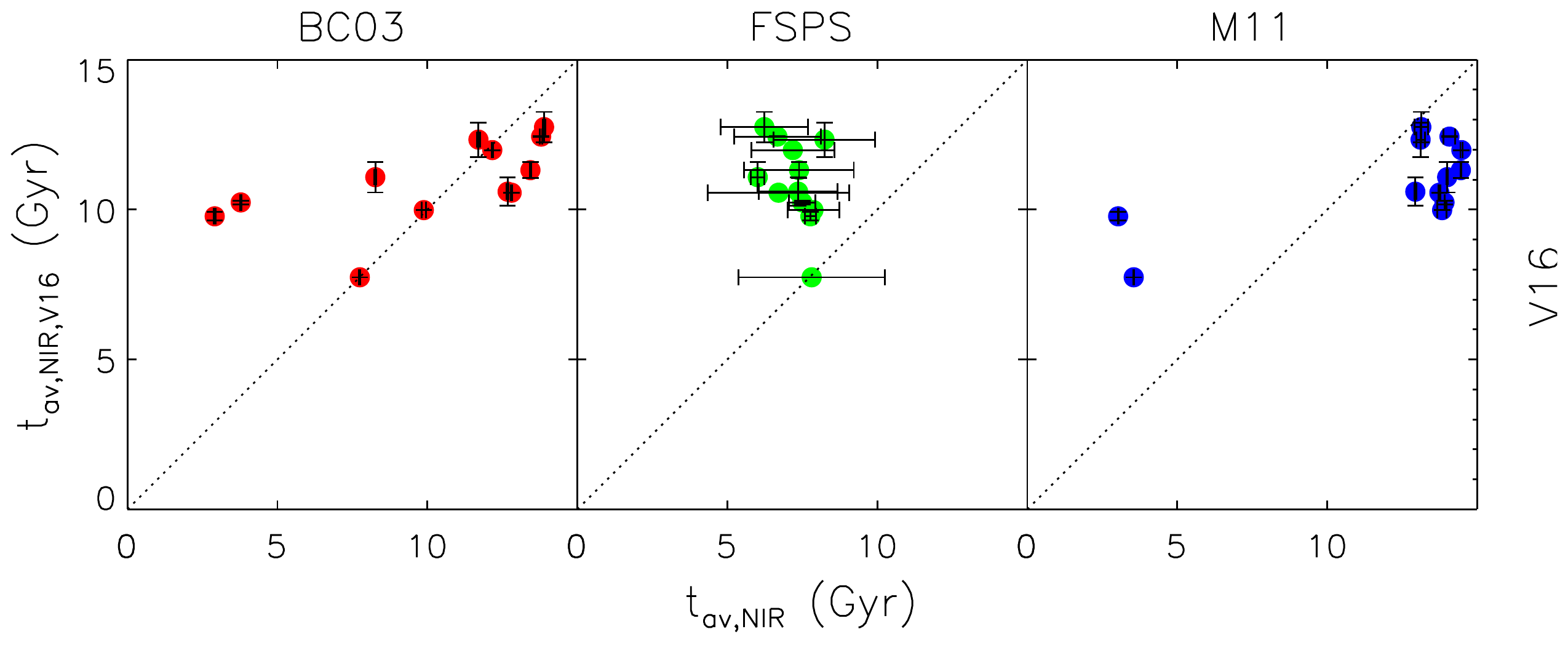}
 	\caption{Mass-weighted mean ages derived from optical (top) and NIR (bottom) spectroscopy for the various models plotted against the ages derived by V16, using a Calzetti extinction curve.}
 \label{fig:meanageredref}
\end{figure}

In \Cref{fig:meanageredref} we plot the mass-weighted mean ages derived by BC03, FSPS and M11 against the mass-weighted mean ages derived by V16. These figures illustrate that all models give consistent results in the optical, and yet differ substantially in the near-infrared. Why is this the case? At optical wavelengths, the change due to age is strong, with H$\beta$ showing a linear decrease as a function of log age (the change in H$\beta$ over the relevant age range is a factor of $\approx25$ greater than the typical measurement error). 
As shown in \Cref{fig:ls,fig:spec}, the near-infrared shows no such strong age variation. Rather, the response to age is very subtle. 
CN1.1 is predicted by the M11 models to vary by a factor 50 greater than the typical  measurement error over the entire range of interest. However, almost all of this variation is at ages $t<3$~Gyr. Looking only at ages $t\geq3$~Gyr, differences are within the errors ($\delta \text{EW}=0.002$~mag vs measurement error = 0.003~mag). Other models display even milder variation.
For CO, the predicted variation is $\approx3-18\times$ the measurement error, depending on the model.
The consequence of this generally subtle variation with age is that the derived properties become driven by systematic effects, with the largest effect being due to the choice of spectral library, as is seen in \Cref{fig:libs,fig:tav_fsps_red}. The choice of isochrones has much less effect, but is still not-negligible. Furthermore, the  subtle age information is likely encoded in the continuum shape, rather than the strengths of specific features, which are observed not to vary strongly across the sample. This means that the common technique of using a high order polynomial to remove the continuum when deriving SFHs gives unreliable results in this wavelength range.

\subsection{Can we constrain treatment of the TP-AGB?}

The lack of age variation in the molecular absorption measured in the data implies that either the TP-AGB phase has less effect on the integrated light of galaxies than is indicated by the M11 models, or that TP-AGB stars are not present in large numbers in any galaxies in our sample. 
We are, of course, dealing with composite populations, rather than SSPs, meaning that the effect of the TP-AGB phase should be diluted in the data compared with the models. 
The question is whether this alone could account for the observed lack of variation in TP-AGB sensitive features despite the broad range in the age-sensitive H$\beta$ index.

We test this using using the M11 models to construct simple two-component mock populations, where the age of the old population is 15~Gyr and the age of the young population varies from $0.3-3$~Gyr. We vary the fraction of young stars from $0-100\%$ and study the effect on the strength of CN1.1 and H$\beta$.

\begin{figure}
  \centering
  	\includegraphics[width=\columnwidth]{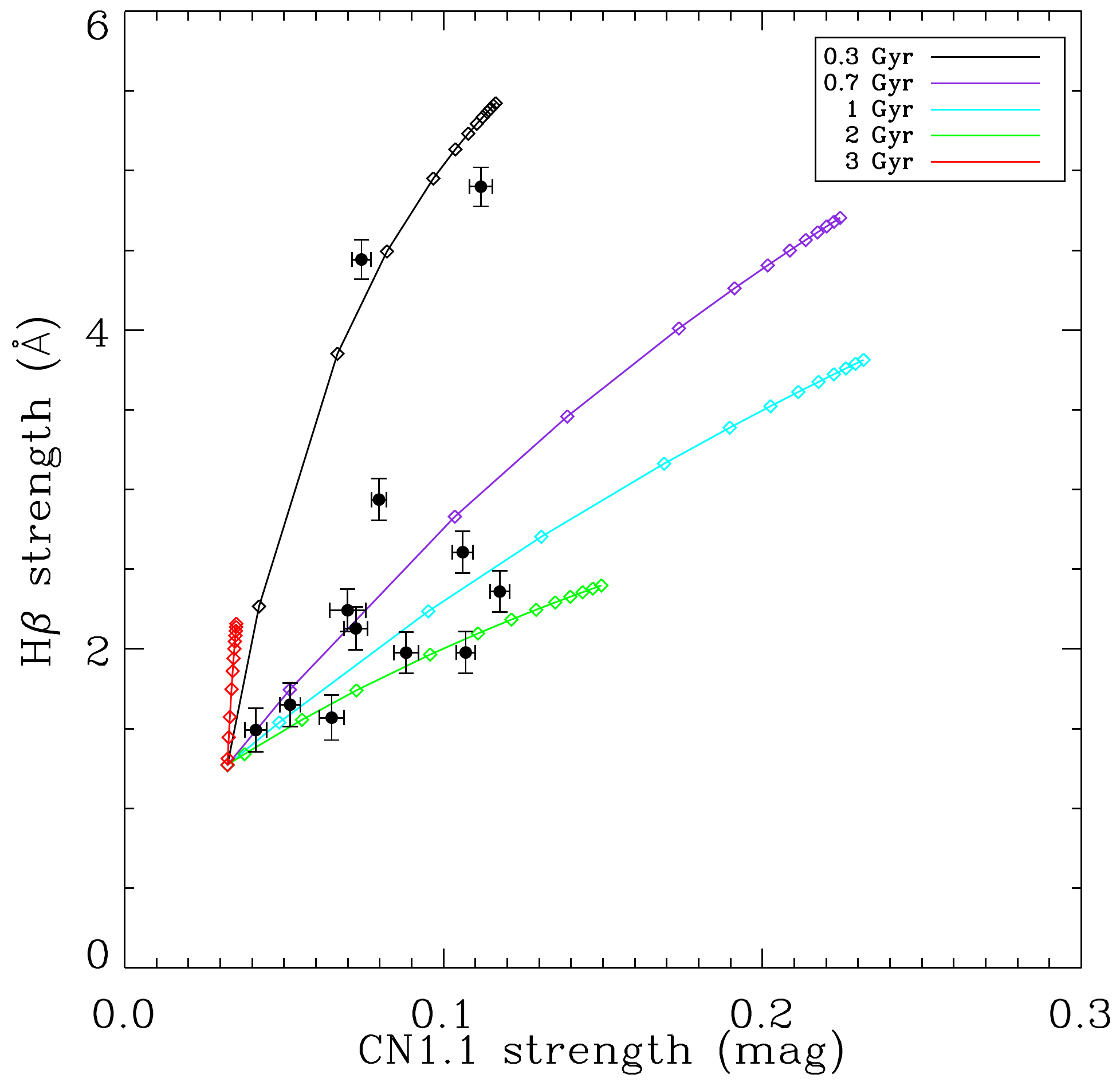}
 	\caption{Toy model showing a comparison between the observed strengths of CN1.1 and H$\beta$ in our sample and the M11 composite models. The coloured lines show composite populations with a young population of various ages, as given in the legend, added onto a population of age 15~Gyr. The fraction of the young population increases along the tracks from left to right, and values are [0, 0.01, 0.05, 0.1, 0.2, 0.3, 0.4, 0.5, 0.6, 0.7, 0.8, 0.9, 1]
.}
\label{fig:toymod}
\end{figure}

\Cref{fig:toymod} shows the strengths of H$\beta$ and CN1.1 measured in our data as filled circles, while the predictions of the various M11 composite populations are shown as coloured tracks. The mass fraction of the young population varies along the tracks. 
Different colours show composite populations with different ages for the young population.
The M11 composite models can reproduce all the indices measured in the sample. 
The very strong H$\beta$ values but only moderate CN values seen in NGC\,3032 and NGC\,3156 can be reproduced by the M11 models assuming a  0.3~Gyr burst making up $\sim 20\%$ of the mass, on top of a dominant, old population. 
The H$\beta$ and CN values measured in the rest of the sample are also able to be reproduced in this scenario. 
In other words, the observed large range of measured H$\beta$ strength 
but only mild variation in CN1.1 strength may be explained by the galaxies being composite populations, and 
thus, we cannot strongly exclude strong TP-AGB contributions to younger populations in our sample without better constraining the shape of the underlying SFH.

Regardless, treatment of the TP-AGB phase does not have a substantial effect on our derived properties. The M11 models, which are unique in their treatment of the TP-AGB phase, give fits which are equally as good as other models (M11 \Chi\/ values are  approximately the same as BC03, which uses the same spectral library), and give SFHs which are equally as good as other models. The M11 and BC03 SFHs are almost identical for most galaxies, again indicating that the derived SFH is largely driven by spectral library rather than specifics of the inclusion of the TP-AGB phase for our particular sample. The situation may be different at high redshift,  because the fraction of stars around the peak TP-AGB period is significantly higher. (see, for example, \cite{kriek_spectral_2010,capozzi_revisiting_2016}). 
As the metallicity dependance of TP-AGB features is not well-known \citep{maraston_evolutionary_2005}, different results may also be obtained for non-solar metallicities.

\subsection{Comparison with previous studies}
This result is in contrast to some other studies, which have found TP-AGB treatment has a large effect on derived properties, with use of the Maraston models being either inconsistent with observations \citep{conroy_propagation_2010,kriek_spectral_2010,zibetti_near-infrared_2013} or more consistent than TP-AGB light models \citep{maraston_evidence_2006,macarthur_integrated_2010, riffel_stellar_2015,capozzi_revisiting_2016}. Most studies to date have been based on photometry, and as such are not directly comparable, but two recent studies have been based on near-infrared spectroscopy - those of \citet[][hereafter Z13]{zibetti_near-infrared_2013}, and \citet[][hereafter R15]{riffel_stellar_2015}.

R15 suggests TP-AGB stars contribute noticeably to a stacked near-infrared spectrum made up largely of Seyfert galaxies from the Palomar survey. This result was based on fitting near-infrared spectra of individual IRTF stars to the stacked galaxy spectrum. However, the authors also find that other evolved stars can reproduce most of the observed absorption features, and the intermediate age galaxies in their sample do not display stronger `TP-AGB features' than the old galaxies. 
This is consistent with what we find in our sample.

Our results seem to contradict Z13, who used optical and near-infrared spectrophotometry of post-starburst galaxies to study the contribution of the TP-AGB phase in galaxies for which it should be at a maximum. The authors looked for the strong molecular features at 1.4 and 1.8~\micron, predicted by the Maraston models to be present at intermediate ages. They did not observe these strong features, which is consistent with what we find (see \Cref{fig:spec} and \Cref{fig:ls}). 
However, they also found that the intermediate age Maraston models, which reproduce the optical age-sensitive indices measured in their sample, produce near-infrared colours which are too red compared to the data. They thus conclude that the Maraston models have too strong a TP-AGB	contribution, as they cannot reproduce the spectrophotometry of post-starburst galaxies under a range of simple or two-component composite population assumptions.

We have two galaxies in our sample which are very similar to the post-starburst galaxies targeted by Z13 (NGC\,3032 and NGC\,3156). Both 
have strong Balmer lines but little or no [OII] or H$\alpha$ emission
and have light-weighted mean ages between 0.5-1.5 Gyr, as calculated from their optical spectra. 
NGC3156 also has SDSS spectra available, which overlaps with the GNIRS wavelength range, meaning that we are able to create a spectrum spanning the entire optical-NIR range, thus avoiding issues of absolutely flux calibrating data from two different instruments.
The SDSS spectrum covers the wavelength range 3808--9208~\AA, while our GNIRS spectrum spans 8500--25,300~\AA. We normalise the spectra in the overlapping wavelength range and stitch the GNIRS spectrum on to the end of the SDSS one. The SDSS spectrum was obtained in a 3'' fibre, centred on the galaxy nucleus, while GNIRS uses a 7x0.3'' slit, from which we extract a spectrum in an $R_{\rm e}$/8 aperture, meaning that the apertures are not exactly matched, however both spectra are dominated by the bright galaxy nucleus.
This situation is similar to the Z13 study, which also compares 3'' SDSS spectra to slit spectra taken with ISAAC.

\begin{figure*}
  \centering
  	\includegraphics[width=\textwidth]{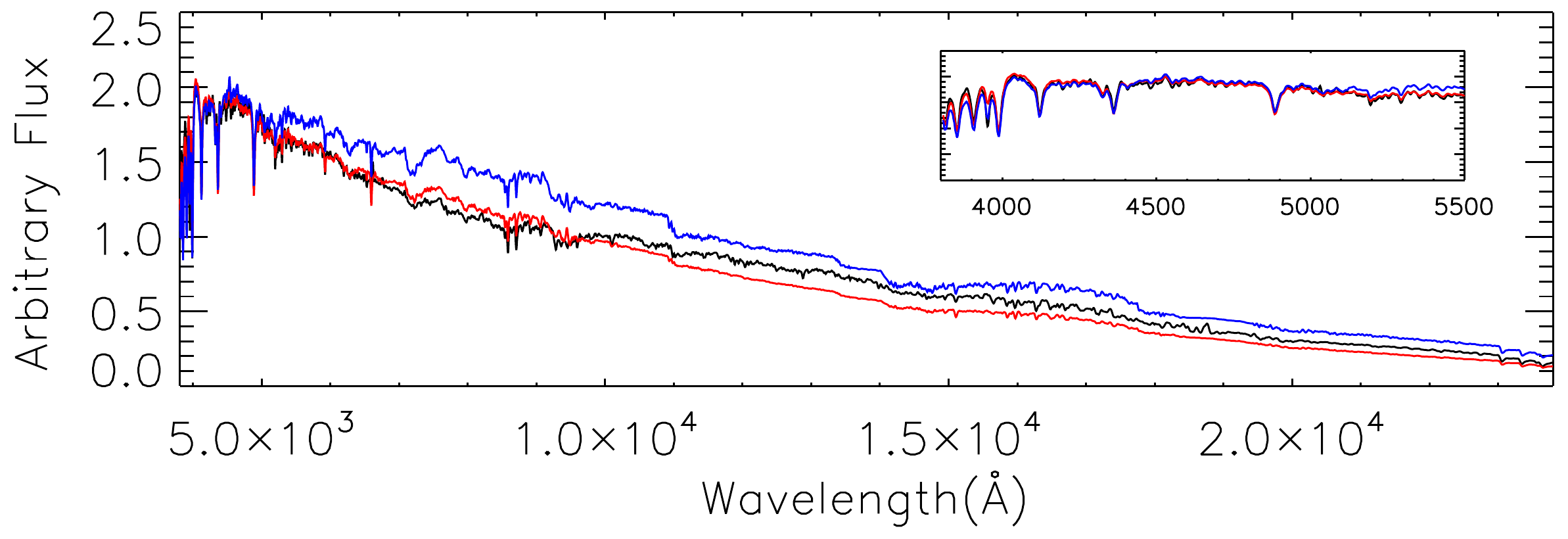}
 	\caption{Results of the fit to post-starburst NGC3156's SDSS+GNIRS spectrum when assuming a two-burst population composed of a 10~Gyr population plus a burst making up 20\% of the mass. The ages of the young populations are 0.3~Gyr (red) and 0.5~Gyr (blue). Both provide excellent fits to the optical, as shown in the inset.}
\label{fig:zibcomp}
\end{figure*}

With our continuous optical-IR spectrum of a post-starburst galaxy, we tested whether we could reproduce the Z13 result. 
First we carried out fits using the composite populations tested by Z13. We constructed simple (two-burst) composite populations, consisting of an old, 10~Gyr component plus a burst. The age of the burst varied, starting from 50~Myr, and the burst fraction varied from 10\% to 100\%. We used the same ages as in Z13, (50~Myr, 0.3, 0.5, 1, 1.5, 2, 10~Gyr). Similarly to Z13, we fit the five optical indices from Gallazzi et al. (2005; D4000$_n$, H$\delta_{\rm{A}} + \rm{H}\gamma_A$, H$\beta$, [MgFe]' and [Mg$_2$Fe])
and extrapolated the fits to the near-infrared. 
Using these simple composite population templates we were unable to fit the entire spectrum well with any combination tested by Z13.
\Cref{fig:zibcomp} shows two of the resulting best fitting composite templates.
We see the same behaviour as is shown in Figure 10 of Z13, which is that a good fit to the optical can overpredict (or underpredict) the near-infrared flux when fitting a simple two-burst population.

\begin{figure*}
  \centering
  	\includegraphics[width=\textwidth]{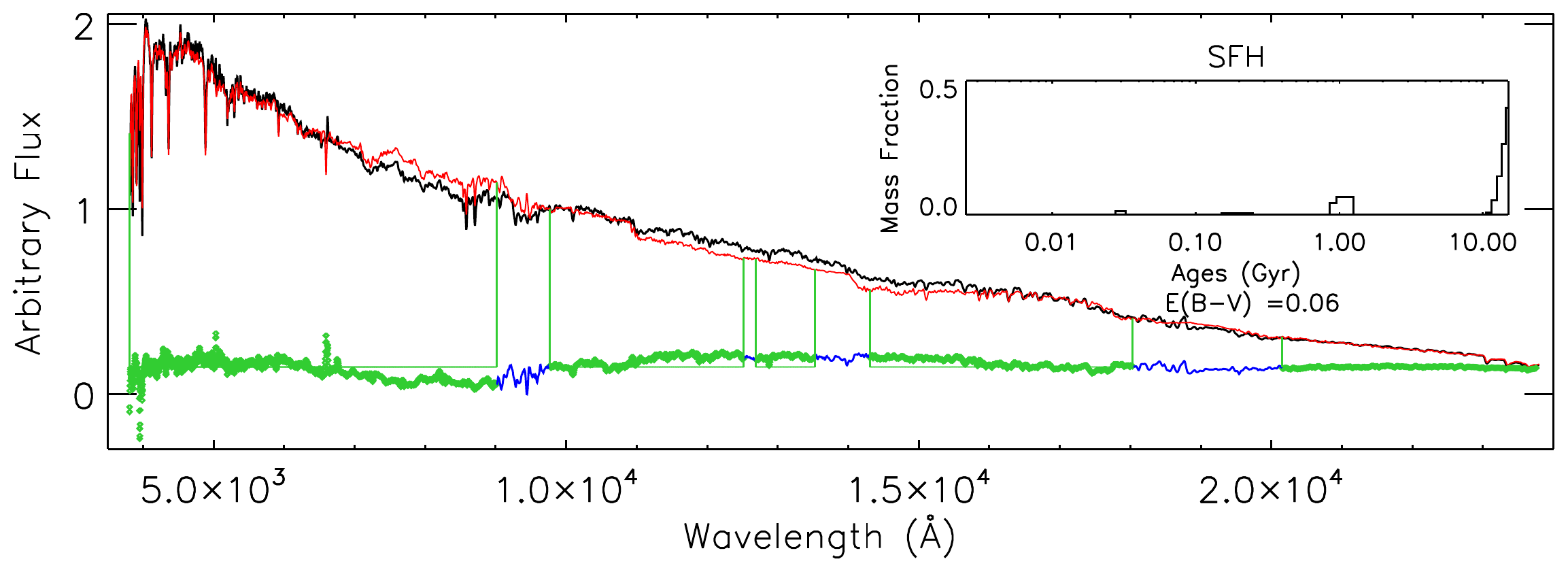}
 	\caption{Results of the fit to post-starburst NGC3156's SDSS+GNIRS spectrum. The pPXF fit is shown in red, and the derived star formation history from this fit is shown in the inset.  }
\label{fig:sdssgnirs}
\end{figure*}

However, this problem is alleviated with our technique of fitting the entire spectral range and allowing for a more general SFH. When we carry out full spectral fitting using multiple templates spanning a large range of ages (\Cref{fig:sdssgnirs}), we see that the M11 models are able to simultaneously match the strong Balmer absorption seen in the optical part of the spectrum, while not overpredicting the near-infrared flux. 
We fit the spectrum using a Calzetti extinction curve, which provides some low-order correction for slight flux calibration errors, but cannot 
correct for any near-infrared colour excess which may be present in the Maraston models - in fact it can only make the spectrum redder.
Directly fitting the spectrum with templates spanning a range of ages 
rather than trying to match individual indices using simple stellar population models or simple two-component populations allows us to better constrain the stellar populations present, and in this case, we find again that the Maraston models are able to match the data with a plausible SFH. 
We therefore interpret the apparent differences between our findings and those of Z13 as deriving from 
our approach of fitting the full spectrum with a more comprehensive star formation history.

\section{Conclusion}

We presented in this paper high quality near-infrared spectra of a sample of galaxies from the \atlas survey. These galaxies were chosen to span a large range of optically-derived ages, from post-starburst galaxies to galaxies with ages close to the age of the universe. This sample was chosen in order to study the effect of age on near-infrared wavelengths, which is currently poorly understood, as various stellar population models make conflicting predictions in this wavelength range.

We fit linear combinations of SSP templates to our data, to derive star formation histories using four different stellar population models, namely \cite{bruzual_stellar_2003}, Conroy, Gunn \& White (2009; 2010), \cite{maraston_stellar_2011} and \cite{vazdekis_uv-extended_2016}. 
We find the V16 models typically give the best fits, owing to their high quality near-infrared stellar spectral libraries. 
We find that derived star formation histories typically differ between the various models, but also depend on the fitting method chosen, even within the same model set. 
Star formation histories derived using an extinction curve to model the continuum are typically more consistent with SFHs derived from optical spectroscopy. Use of a multiplicative polynomial to model the continuum causes models to lose age sensitivity, indicating that in the near-infrared regime, 
age sensitive information is encoded largely in the continuum shape rather than in the strengths of specific features. This is supported by our measurement of the strengths of specific indices, most of which do not vary strongly as a function of optical age. 

We find that the largest differences in derived SFHs are caused by the  choice of stellar spectral library. While the low-resolution near-infrared BaSeL library in FSPS results in poor agreement between optical and near-infrared derived SFHs, the use of the high resolution theoretical C3K library results 
in the most self-consistent SFHs across the two wavelength ranges. We therefore suggest that the inclusion of high quality near-infrared stellar spectral libraries (such as IRTF, XSL, C3K) into current near-infrared stellar population models should be a top priority for modellers.

Inclusion of the TP-AGB phase appears to have only a second-order effect in this sample, with ``TP-AGB heavy" and ``TP-AGB light" models resulting in comparable fits and SFHs.
In particular, the M11 and BC03 models, when using the same spectral library, produce very similar SFHs from full spectral fitting, despite a very different TP-AGB treatment. 
We do not observe strong variation in TP-AGB sensitive indices as a function of age, which is predicted by the single-burst M11 models; however optical and near-infrared indices H$\beta$ and CN can be reproduced by a two-burst population constructed with M11, including TP-AGB stars, so our galaxies cannot strongly constrain the contribution of the TP-AGB. 
Finally, we compare to previous works addressing the TP-AGB fraction in galaxies. We find general agreement with our findings where comparable data were used, and apparent disagreements with studies using lower quality data may be explained through the combination of higher-quality data and the use of a more general spectral fitting approach. This demonstrates the importance of applying this type of analysis to high quality data with a large, continuous wavelength coverage for a range of galaxy types to provide more stringent tests on stellar population models in the future.

\section*{Acknowledgments}
We thank the anonymous referee for their constructive review, which helped improve the paper. We thank Rachel Mason for making the XDGNIRS reduction pipeline available. CMB acknowledges support from the Australian Government Research Training Program Scholarship.
Based on observations obtained at the Gemini Observatory, which is operated by the Association of Universities for Research in Astronomy, Inc., under a cooperative agreement with the NSF on behalf of the Gemini partnership: the National Science Foundation (United States), the National Research Council (Canada), CONICYT (Chile), Ministerio de Ciencia, Tecnolog\'{i}a e Innovaci\'{o}n Productiva (Argentina), and Minist\'{e}rio da Ci\^{e}ncia, Tecnologia e Inova\c{c}\~{a}o (Brazil). 
RMcD is the recipient of an Australian Research Council Future Fellowship (project number FT150100333).

\bibliographystyle{mn2e}
\bibliography{bib3}

\appendix
\section{Tables}


\begin{table*}
\centering 
\caption{Results for the near-infrared fits using a Calzetti extinction curve. }\begin{tabular}{l c c c c c c c c c c c c c c c} \hline
\multicolumn{16}{c}{\vspace{-0.3cm}} \\ 
&\multicolumn{3}{c}{BC} &   &   \multicolumn{3}{c}{FSPS}  & &   \multicolumn{3}{c}{V16}   &  & \multicolumn{3}{c}{M11}\\ 
Galaxy & \Chi & $$E(B-V)$$ &  $t_{{\rm av}}$ &  &     \Chi & $$E(B-V)$$&  $t_{{\rm av}}$  &   &    \Chi &$$E(B-V)$$ & $t_{{\rm av}}$   &  &     \Chi  &   $$E(B-V)$$ &  $t_{{\rm av}}$ \\  
\cline{2-4}\cline{6-8}\cline{10-12}\cline{14-16}
IC0719&12.74& 0.07& 9.88&& 68.91& 0.20& 7.87&& 5.06& 0.10& 9.98&&14.19& 0.16&13.83\\
NGC3032&12.33& 0.00& 2.91&& 40.99& 0.04& 7.77&& 2.72& 0.00& 9.78&& 6.42& 0.05& 3.03\\
NGC3098&10.20& 0.06&11.70&& 47.10& 0.16& 8.24&& 2.64& 0.08&12.33&& 7.91& 0.13&13.11\\
NGC3156&13.51& 0.00& 3.78&& 50.17& 0.05& 7.50&& 3.79& 0.00&10.24&&11.45& 0.00&13.91\\
NGC3182& 9.46& 0.05&13.79&& 74.44& 0.19& 6.67&& 4.44& 0.09&12.43&&12.23& 0.15&14.08\\
NGC3301&15.47& 0.00&13.89&& 74.75& 0.12& 6.23&& 5.49& 0.03&12.75&&15.58& 0.08&13.14\\
NGC3489&23.50& 0.00&12.81&&127.67& 0.11& 6.71&&11.68& 0.03&10.56&&26.13& 0.08&13.74\\
NGC4379&20.29& 0.04&12.16&&168.35& 0.18& 7.18&& 8.99& 0.09&11.98&&26.00& 0.14&14.47\\
NGC4578&14.11& 0.04&13.44&&185.24& 0.17& 7.39&& 5.94& 0.08&11.31&&20.40& 0.13&14.46\\
NGC4608&10.84& 0.02& 8.26&& 89.35& 0.14& 6.01&& 2.77& 0.04&11.08&&11.86& 0.09&14.00\\
NGC4710&14.38& 0.53& 7.75&& 18.64& 0.57& 7.81&& 2.63& 0.54& 7.74&& 7.06& 0.57& 3.55\\
NGC5475&15.04& 0.00&12.67&& 52.36& 0.12& 7.36&& 3.05& 0.03&10.60&&11.48& 0.09&12.94\\\cline{2-4}\cline{6-8}\cline{10-12}\cline{14-16}
\hline
\end{tabular}
\label{tab:fittablered}
\end{table*} 

\begin{table*}
\centering 
\caption{Results for the optical fits using a Calzetti extinction curve. }\begin{tabular}{l c c c c c c c c c c c c c c c} \hline
\multicolumn{16}{c}{\vspace{-0.3cm}} \\ 
&\multicolumn{3}{c}{BC} &   &   \multicolumn{3}{c}{FSPS}  & &   \multicolumn{3}{c}{V16}   &  & \multicolumn{3}{c}{M11}\\ 
Galaxy & \Chi & $$E(B-V)$$ &  $t_{{\rm av}}$ &  &     \Chi & $$E(B-V)$$&  $t_{{\rm av}}$  &   &    \Chi &$E(B-V)$ & $t_{{\rm av}}$   &  &     \Chi  &   $E(B-V)$ &  $t_{{\rm av}}$ \\  
\cline{2-4}\cline{6-8}\cline{10-12}\cline{14-16}

IC0719& 7.39& 0.46&12.35&&  3.42& 0.51& 9.03&& 4.21& 0.48&10.55&& 3.09& 0.45&12.53\\
NGC3032& 4.68& 0.21& 1.00&&  5.56& 0.24& 1.00&& 4.56& 0.22& 1.00&& 7.47& 0.09& 1.00\\
NGC3098& 8.26& 0.36& 9.15&&  3.93& 0.39& 6.30&& 5.14& 0.37& 7.33&& 3.74& 0.34&11.16\\
NGC3156& 4.04& 0.31& 1.01&&  2.80& 0.34& 1.12&& 2.59& 0.32& 1.12&& 7.26& 0.19& 1.00\\
NGC3182&15.39& 0.14&13.86&&  6.67& 0.22&12.05&& 8.11& 0.17&12.75&& 7.94& 0.15&13.83\\
NGC3301&11.12& 0.33& 6.77&&  4.67& 0.36& 5.23&& 6.99& 0.34& 6.19&& 3.75& 0.28& 7.97\\
NGC3489& 7.88& 0.29&12.57&&  2.79& 0.32& 8.54&& 3.63& 0.30&10.41&& 2.96& 0.24& 8.75\\
NGC4379&12.71& 0.29&13.46&&  4.67& 0.35&10.96&& 6.50& 0.32&12.40&& 5.76& 0.31&13.63\\
NGC4578&26.15& 0.21&13.98&& 10.03& 0.29&12.46&&13.34& 0.25&13.01&&12.96& 0.24&14.02\\
NGC4608&26.63& 0.22&13.76&&  7.97& 0.28&11.31&&12.21& 0.25&12.77&&10.33& 0.24&13.99\\
NGC4710&11.74& 0.53& 9.03&&  6.53& 0.56& 6.47&& 8.29& 0.54& 7.22&& 6.08& 0.49&10.88\\
NGC5475& 5.83& 0.24&12.93&&  3.13& 0.27& 7.21&& 2.78& 0.25&10.26&& 2.61& 0.22&11.88\\
\cline{2-4}\cline{6-8}\cline{10-12}\cline{14-16}\hline
\end{tabular}
\label{tab:fittableredopt}
\end{table*} 


\begin{table*}
\centering 
\caption{Results for the near-infrared fits using an order 10 multiplicative polynomial. }
\begin{tabular}{l c c c c c c c c c c c c c c c} \hline
\multicolumn{12}{c}{\vspace{-0.3cm}} \\ 
&\multicolumn{2}{c}{BC03} &   &   \multicolumn{2}{c}{FSPS}  & &   \multicolumn{2}{c}{V16}   &  & \multicolumn{2}{c}{M11}\\ 
Galaxy & \Chi &  $t_{{\rm av}}$ &  &     \Chi & $t_{{\rm av}}$  &   &    \Chi & $t_{{\rm av}}$   &  &   \Chi  &   $t_{{\rm av}}$ \\  
\hline
IC0719& 6.24& 6.51&& 7.13& 3.69&& 1.96&12.66&& 4.53& 5.52\\
NGC3032& 4.03& 4.02&& 3.57&12.99&& 1.11& 8.24&& 2.76& 3.04\\
NGC3098& 4.36& 4.19&& 3.91&12.25&& 1.23&12.08&& 2.76& 4.78\\
NGC3156& 4.31& 3.93&& 3.68&10.69&& 1.41& 8.62&& 3.12& 4.69\\
NGC3182& 3.18& 3.72&& 3.88&12.38&& 1.22&11.50&& 2.32& 4.14\\
NGC3301& 5.07& 3.82&& 4.44&10.21&& 1.46&12.27&& 3.11& 4.76\\
NGC3489& 7.29& 4.59&& 5.34&10.08&& 2.17&11.01&& 5.03& 4.32\\
NGC4379& 4.95& 4.26&& 3.98&10.00&& 1.48&12.67&& 3.60& 4.83\\
NGC4578& 5.28& 3.02&& 5.70&12.88&& 1.54&11.55&& 4.32& 3.65\\
NGC4608& 4.71& 2.74&& 4.08&13.26&& 1.15& 9.71&& 3.82& 3.55\\
NGC4710& 5.11& 4.48&& 4.02& 4.91&& 1.46& 9.57&& 3.57& 3.76\\
NGC5475& 5.95& 4.77&& 4.60& 8.76&& 1.48&11.98&& 3.88& 3.16\\
\hline
\end{tabular}
\label{tab:fittablemdeg}
\end{table*} 

\begin{table*}
\centering 
\caption{Results for the optical fits using a multiplicative polynomial. }
\begin{tabular}{l c c c c c c c c c c c c c c c} \hline
\multicolumn{12}{c}{\vspace{-0.3cm}} \\ 
&\multicolumn{2}{c}{BC03} &   &   \multicolumn{2}{c}{FSPS}  & &   \multicolumn{2}{c}{V16}   &  & \multicolumn{2}{c}{M11}\\ 
Galaxy & \Chi &  $t_{{\rm av}}$ &  &     \Chi & $t_{{\rm av}}$  &   &    \Chi & $t_{{\rm av}}$   &  &   \Chi  &   $t_{{\rm av}}$ \\  
\hline
IC0719& 5.57&13.34&& 2.29& 7.14&& 2.20& 9.39&& 2.23&13.00\\
NGC3032& 4.84& 1.00&& 5.49& 1.00&& 4.25& 1.00&& 5.43& 1.00\\
NGC3098& 7.06& 9.79&& 3.00& 4.76&& 3.71& 6.18&& 3.14& 8.69\\
NGC3156& 4.08& 1.00&& 2.57& 1.00&& 1.89& 1.00&& 6.13& 1.00\\
NGC3182&12.49&14.06&& 4.13&12.33&& 4.69&12.99&& 5.22&14.05\\
NGC3301& 9.33& 7.27&& 2.98& 4.10&& 4.30& 5.69&& 3.28& 5.24\\
NGC3489& 6.49&13.09&& 2.23& 7.00&& 1.91& 8.98&& 3.09& 9.37\\
NGC4379&11.47&13.57&& 4.37&10.39&& 5.52&12.05&& 4.48&14.13\\
NGC4578&22.67&14.24&& 6.21&12.73&& 8.74&13.36&& 9.07&14.40\\
NGC4608&24.01&14.02&& 6.80&10.89&& 9.78&12.74&& 7.59&14.34\\
NGC4710& 8.11&10.39&& 4.19& 4.51&& 4.77& 5.97&& 4.48& 6.55\\
NGC5475& 5.20&13.31&& 2.68& 6.28&& 1.99& 8.97&& 2.53&11.21\\
\hline
\end{tabular}
\label{tab:fittablemdegopt}
\end{table*}

\begin{table*}
\centering 
\caption{Results of updated FSPS models with Calzetti extinction curve}
\begin{tabular}{l c c c c c c c c c c c } \hline
\multicolumn{12}{c}{\vspace{-0.3cm}} \\ 
& &  \multicolumn{4}{c}{Padova} & &  &  \multicolumn{4}{c}{MIST} \\
\multicolumn{12}{c}{\vspace{-0.3cm}} \\ 
& &  \multicolumn{2}{c}{Near-IR} && Optical &  &  &  \multicolumn{2}{c}{Near-IR}&  & Optical\\
Galaxy & & \Chi &  $t_{{\rm av}}$ & & $t_{{\rm av}}$ & & &    
 \Chi & $t_{{\rm av}}$ & & $t_{{\rm av}}$   \\  
\hline
IC0719  & & 4.26 & 11.47 &&  9.71& & & 4.25 & 12.97 & & 10.33\\
NGC3032 & & 4.07 & 8.41 & &  1.08 &  &  & 4.15 & 3.72 & &  3.32 \\
NGC3098 && 2.92 & 7.60 &  &  5.57 &  &  &   2.97& 7.67& & 6.21\\
NGC3156 & & 3.49  & 6.48 & & 1.19 &  &  & 3.48&  4.10 &  &  1.50 \\
NGC3182 &  & 4.0 & 12.19 &&  12.00 &  & & 4.05& 13.02&  & 12.18\\
NGC3301  & & 5.95 & 5.68 & & 4.14 & &  & 6.06&   11.18 & & 4.63\\
NGC3489 & & 11.00&  6.65 &  &  6.50 &  & &  11.30 & 11.59 & &  8.73\\
NGC4379 & &    7.32 & 12.00 & &9.43 &  &  & 7.40& 13.33& &  10.69 \\
NGC4578  & & 5.29 &  12.98  &  & 12.08 &  &  & 5.31 & 13.26 &&  12.23\\
NGC4608  & & 2.25&  11.25&   &  11.29&  &  &  2.35& 11.06& &   11.84 \\
NGC4710  &   & 2.91&  4.64&  & 5.57& &  & 2.93&  3.46& &  6.61\\
NGC5475  &   & 3.62 & 6.98& & 9.84  & &  &      3.72&  8.26& & 10.84\\

\hline
\end{tabular}
\label{tab:fspsproperties}
\end{table*}

\begin{table*}
\centering 
\caption{Line strengths measured in the data, at $R\approx500$}
\resizebox{\textwidth}{!}{\begin{tabular}{lccccccccc}\hline
\multicolumn{10}{c}{\vspace{-0.3cm}} \\ 
Galaxy & CN1.1 (mag) & Na1.14 (\AA)& CN1.4 (mag) & C$_2$ (mag) & Na 2.21 (\AA) & $\langle$Fe$\rangle $ (\AA) & Ca (\AA) & Mg (\AA)& CO (mag)\\  \hline 

IC0719&0.072$\pm$0.004&0.943$\pm$0.282&0.097$\pm$0.004&0.053$\pm$0.005&3.115
$\pm$0.277&1.13$\pm$0.238&2.161$\pm$0.582&0.567$\pm$0.3&1.197$\pm$0.004\\
NGC3032&0.112$\pm$0.003&2.254$\pm$0.303&0.12$\pm$0.004&0.049$\pm$0.005&5.056
$\pm$0.307&0.977$\pm$0.272&3.913$\pm$0.618&0.798$\pm$0.334&1.219$\pm$0.005\\
NGC3098&0.107$\pm$0.003&1.189$\pm$0.232&0.094$\pm$0.003&0.04$\pm$0.004&2.82$\pm$
0.241&0.793$\pm$0.206&2.375$\pm$0.457&0.898$\pm$0.262&1.192$\pm$0.004\\
NGC3156&0.074$\pm$0.003&0.798$\pm$0.257&0.103$\pm$0.003&0.042$\pm$0.004&2.559
$\pm$0.257&0.551$\pm$0.22&2.256$\pm$0.498&0.151$\pm$0.266&1.198$\pm$0.004\\
NGC3182&0.088$\pm$0.004&1.323$\pm$0.285&0.094$\pm$0.004&0.041$\pm$0.005&3.347
$\pm$0.284&1.048$\pm$0.248&2.831$\pm$0.566&0.704$\pm$0.298&1.202$\pm$0.005\\
NGC3301&0.106$\pm$0.003&1.066$\pm$0.261&0.11$\pm$0.004&0.041$\pm$0.005&2.797
$\pm$0.278&0.91$\pm$0.224&3.081$\pm$0.54&0.689$\pm$0.278&1.211$\pm$0.004\\
NGC3489&0.08$\pm$0.002&0.722$\pm$0.193&0.121$\pm$0.002&0.047$\pm$0.003&2.929
$\pm$0.196&1.007$\pm$0.164&2.398$\pm$0.403&0.614$\pm$0.211&1.206$\pm$0.003\\
NGC4379&0.041$\pm$0.003&0.364$\pm$0.247&0.099$\pm$0.003&0.052$\pm$0.005&3.589
$\pm$0.262&0.568$\pm$0.225&1.577$\pm$0.49&0.746$\pm$0.302&1.184$\pm$0.004\\
NGC4578&0.052$\pm$0.003&1.253$\pm$0.225&0.101$\pm$0.003&0.05$\pm$0.004&4.121
$\pm$0.234&1.057$\pm$0.197&2.377$\pm$0.449&0.792$\pm$0.247&1.208$\pm$0.004\\
NGC4608&0.065$\pm$0.004&0.838$\pm$0.28&0.105$\pm$0.003&0.056$\pm$0.005&2.972
$\pm$0.296&0.939$\pm$0.237&2.913$\pm$0.547&1.273$\pm$0.289&1.186$\pm$0.004\\
NGC4710&0.07$\pm$0.005&0.555$\pm$0.346&0.062$\pm$0.004&0.032$\pm$0.005&3.234
$\pm$0.358&1.115$\pm$0.282&2.68$\pm$0.641&0.943$\pm$0.362&1.204$\pm$0.005\\
NGC5475&0.118$\pm$0.003&1.045$\pm$0.222&0.116$\pm$0.003&0.052$\pm$0.004&3.742
$\pm$0.233&1.147$\pm$0.185&3.143$\pm$0.442&1.017$\pm$0.237&1.211$\pm$0.003\\

\hline 
\end{tabular}}
\end{table*}

\bsp

\label{lastpage}

\end{document}